\documentclass[times,twocolumn]{aastex631}
\usepackage{amsmath,bm}%,caption}
\usepackage{comment}
\usepackage{mathtools} %loads amsmath as well
\usepackage{enumitem}
\usepackage[caption=false]{subfig}
\usepackage{xcolor, colortbl}
\definecolor{Gray}{gray}{0.93}
\definecolor{DarkerGray}{gray}{0.8}
\newcolumntype{a}{>{\columncolor{Gray}[\tabcolsep]}c}
\newcolumntype{b}{>{\columncolor{white}}c}

\newcommand{\uat}[2]{\href{http://astrothesaurus.org/uat/#1}{#2  (#1)}}

\submitjournal{The Astrophysical Journal}
\shorttitle{CIB Tomography}
\shortauthors{Chiang et al.}
\graphicspath{{./}{figures/}}
\defcitealias{2020ApJ...902...56C}{C20}

\begin{document}

\title{Cosmic Infrared Background Tomography\\and a Census of Cosmic Dust and Star Formation}

\email{ykchiang@asiaa.sinica.edu.tw}

\author[0000-0001-6320-261X]{Yi-Kuan Chiang}
\affiliation{Academia Sinica Institute of Astronomy and Astrophysics (ASIAA), No. 1, Section 4, Roosevelt Road, Taipei 10617, Taiwan}

\author[0000-0001-5133-3655]{Ryu Makiya}
\affiliation{Academia Sinica Institute of Astronomy and Astrophysics (ASIAA), No. 1, Section 4, Roosevelt Road, Taipei 10617, Taiwan}

\author[0000-0003-3164-6974]{Brice M\'enard}
\affiliation{Department of Physics \& Astronomy, Johns Hopkins University, 3400 N. Charles Street, Baltimore, MD 21218, USA}
\affiliation{Santa Fe Institute, 1399 Hyde Park Road, Santa Fe, NM 87501, USA}




\begin{abstract}

The cosmic far-infrared background (CIB) encodes dust emission from all galaxies and carries valuable information on structure formation, star formation, and chemical enrichment across cosmic time. However, its redshift-dependent spectrum remains poorly constrained due to line-of-sight projection effects. We address this by cross-correlating 11 far-infrared intensity maps spanning a 50-fold frequency range from \textit{Planck}, \textit{Herschel}, and \textit{IRAS}, with spectroscopic galaxies and quasars from SDSS I--IV tomographically. We mitigate foregrounds using a CIB-free Milky Way dust map. These cross-correlation amplitudes on two-halo scales trace bias-weighted CIB redshift distributions and collectively yield a $60\sigma$ detection of the evolving CIB spectrum, sampled across hundreds of rest-frame frequencies over $0 < z < 4$. We break the bias--intensity degeneracy by adding monopole information from FIRAS$+$\textit{Planck}. The recovered spectrum reveals a dust temperature distribution that is broad, spanning the full range of host environments, and moderately evolving. Using low-frequency CIB amplitudes, we constrain cosmic dust density, $\Omega_{\rm dust}$, which peaks at $z = 1$--$1.5$ and declines threefold to the present. Our broad spectral coverage enables a determination of the total infrared luminosity density to 0.04~dex statistical precision, tracing star-formation history with negligible cosmic variance across 90\% of cosmic time. We find that cosmic star formation is 80\% dust-obscured at $z = 0$ and 60\% at $z = 4$. Our results, based on intensity mapping, are complete, requiring no extrapolation to faint galaxies or low-surface-brightness components. We release our tomographic CIB spectrum and redshift distributions as a public resource for future studies of the CIB, both as a cosmological matter tracer and CMB foreground.

\end{abstract}


\section{Introduction} \label{sec:intro}
The cosmic far-infrared background (CIB) is the total far-infrared emission produced throughout the universe during the epoch of galaxy formation. It serves as a key cosmological probe for quantifying the net efficiency of galaxy assembly, cosmic star formation, and the life cycle of cosmic dust, all ultimately driven by the gravitational growth of structures in the universe \citep{1969RSPTA.264..279P, 1986ApJ...306..428B}.

The CIB is most commonly observed as an integrated radiation field, which could be resolved into individual galaxies in only limited areas with deep and high-resolution data. As part of the extragalactic background light (EBL), the CIB carries a significant energy share compared to other wavelengths \citep{1998ApJ...508...25H, 1999A&A...344..322L}. It exhibits a thermal spectrum peaking at 100--$200\,\mu$m, as dust in the interstellar medium (ISM) absorbs UV photons from young stars and re-emits in the far-infrared. These absorbing and emitting grains consist of carbon and silicate dust mixtures, ranging in size from a few angstroms to a few microns \citep{2003ARA&A..41..241D}. In addition to star formation, active galactic nuclei (AGNs) heat dust in their surrounding medium, while their contribution is only significant in the mid-infrared \citep{2001ARA&A..39..249H}. Given this emission mechanism, detailed characterizations of the CIB are crucial for understanding the formation of stars and dust, two key baryonic products of galaxy formation in the cosmic inventory \citep{1998ApJ...508..106D, 2004ApJ...616..643F, 2005ARA&A..43..727L,2005PhR...409..361K}. 

The CIB also plays a significant role in modern cosmology. Its strength, relatively high bias factor, and high redshift make it a unique tracer of large-scale structure (LSS) for clustering and gravitational lensing studies, often via cross-correlations with other datasets \citep[][]{2013JCAP...05..004H,2014A&A...571A..18P,2014A&A...570A..98S,2015MNRAS.446.2696S,2016MNRAS.463.2046T,2018PhRvD..97l3539S,2022A&A...665A..52Y,2024JCAP...05..058Y}. However, its spatial and spectral fluctuations introduce systematics in cosmic microwave background (CMB) experiments, affecting both primary CMB analyses and secondary signals such as the Sunyaev-Zeldovich (SZ) effect and other types of spectral distortions \citep{1972CoASP...4..173S, 2014A&A...571A..15P,2014A&A...571A..21P,2016SPIE.9904E..0WK,2019BAAS...51c.184C,2020ApJ...902...56C,2021ApJ...914...68Z}. The CIB also contaminates infrared-based Milky Way dust reddening maps \citep[e.g.,][]{1998ApJ...500..525S, 2014A&A...571A..11P}, which can impact supernova cosmology, weak lensing, and galaxy clustering through correlated biases in extinction correction \citep{2019ApJ...870..120C}.

One key challenge in studying the CIB is the projection effect over a broad redshift kernel, roughly $z\sim1$--$4$. Over this extended timespan, both the amplitude and spectral shape of the CIB must have evolved alongside the underlying galaxy populations. Any astrophysical or cosmological application involving the CIB, therefore, depends on assumptions about a fundamental but uncertain property---the redshift-dependent spectrum or spectral energy distribution (SED). This can be characterized through the comoving emissivity $\epsilon_{\nu}$ as a function of rest-frame frequency $\nu$ and redshift $z$. A fully equivalent description is to slice the CIB by observed frequency and examine the redshift distributions, which vary strongly with frequency due to different ``$K$-corrections.''

The lack of a detailed CIB spectrum has posed significant challenges. For instance, uncertainties in the peak frequency introduce large systematics in cosmic star formation estimates \citep[][]{2013A&A...557A..66B, 2013ApJ...768...58T,  2018A&A...614A..39M, 2022A&A...665A..52Y}, as the scaling between total infrared luminosity $L_{\rm TIR}$ and dust temperature $T$ is steep, approximately $L_{\rm TIR} \propto T^{6}$ \citep[][]{2016MNRAS.461.1328E}. At low frequencies, assumptions about the spectral shape and its evolution, which depend on the optical properties of dust, can impact the construction of Compton $y$ maps for the SZ effect \citep[][]{2024PhRvD.109f3530C,2024PhRvD.109b3528M}. Lastly, there is a large body of literature on projected CIB power spectra, while the physical implications depend heavily on the assumed SED and redshift distributions, both typically derived from models that lack direct observational validation \citep{
2009ApJ...707.1766V, 2011A&A...536A..18P, 2012ApJ...755...70R, 2012A&A...537A.137P, 2012MNRAS.421.2832S, 2013ApJ...772...77V, 2013MNRAS.436.1896A, 2014A&A...571A..30P, 2015ApJ...799..177G, 2017MNRAS.466.4651W, 2018A&A...614A..39M, 2019A&A...621A..32M, 2020MNRAS.491.1079R,2023MNRAS.520.1895J}. 

In this paper, we present a precision measurement of the CIB spectrum $\epsilon_{\nu}(\nu, z)$ across a 50-fold range in frequency (observed frame $\rm 100\,GHz$ to $\rm 5\,THz$) and $90\%$ of cosmic time over $0<z<4$. We tomographically cross-correlate CIB anisotropies in intensity maps from the {\it Planck} satellite, {\it Herschel} Space Observatory, and the Infrared Astronomical Satellite ({\it IRAS}) with spectroscopic reference galaxies and quasars (QSOs) from the Sloan Digital Sky Survey (SDSS) as a function of redshift. This approach follows clustering-based redshift estimation \citep{2008ApJ...684...88N,2013MNRAS.433.2857M,2013arXiv1303.4722M}, applied to diffuse CIB data in a manner similar to the pioneering analysis by \cite{2015MNRAS.446.2696S}. Using the generic EBL tomography formalism from \cite{2019ApJ...877..150C}, we combine multi-band redshift distributions to construct a CIB spectrum densely sampled in both redshift and rest-frame frequency.

An earlier measurement of our EBL tomography, using eight bands from {\it Planck} and {\it IRAS}, was presented in \citet[][hereafter \citetalias{2020ApJ...902...56C}]{2020ApJ...902...56C}, where the focus was on the cosmic thermal history of gas, examined through the SZ effect superimposed on the CIB at low frequencies. In a second paper, \cite{2021ApJ...910...32C}, we provided a theoretical interpretation linking this thermal history to gravitational structure formation. In this third paper of the series, we shift the focus to the CIB by adding three more bands from {\it Herschel} near the peak frequency. We also improve foreground mitigation using the new CIB-free dust map from \cite{2023ApJ...958..118C} before performing the cross-correlation measurements.

This paper is organized as follows. Section~\ref{sec:data} describes the data processing. Section~\ref{sec:result} introduces the clustering redshift formalism and presents our main result, the tomographic CIB spectrum. In Section~\ref{sec:interpretation}, we develop a physical yet flexible CIB SED model to extract information on dust temperature, mass, and cosmic star formation. We discuss broader implications in Section~\ref{sec:discussion} and summarize in Section~\ref{sec:summary}. Throughout, we adopt the {\it Planck} 2018 cosmology \citep{2020A&A...641A...6P} with ($h$, $\Omega_{\rm c}h^2$, $\Omega_{\rm b}h^2$, $A_{\rm s}$, $n_{\rm s}$) = (0.6737, 0.1198, 0.02233, $2.097\times10^{-9}$, 0.9652).

\section{Data}\label{sec:data}

\subsection{Intensity Maps in Planck, Herschel, and IRAS}

We begin with the same eight full-sky intensity maps used in \citetalias{2020ApJ...902...56C}, which include six channels from the High Frequency Instrument (HFI) of {\it Planck} at 100, 143, 217, 353, 545, and 857 GHz from \cite{2016A&A...594A...8P}, along with the 100 and $60\ \mu$m (3 and 5 THz) bands from {\it IRAS}, using the re-calibrated product from \cite{2005ApJS..157..302M}. The beam full width at half maximum (FWHM) is $10'$, $7'.1$, $5'.5$, $5'$, $5'$, $5'$, $4'.3$, and $4'$ from low to high frequencies, respectively. The maps are sampled using the HEALPix \citep{2005ApJ...622..759G} scheme at $N_{\rm side}=2048$ resolution and are provided in specific intensity $I_{\nu}$ in $\rm MJy\ sr^{-1}$, assuming an in-band spectrum following the $IRAS$ convention of $\nu I_{\nu}= \mathrm{constant}$. The four maps up to 353 GHz were cleaned of primary CMB contamination in \citetalias{2020ApJ...902...56C} using the \cite{2016A&A...591A..50B} CMB map, reducing noise in the subsequent CIB analyses. Our CIB results are not sensitive to the choice of CMB template: while we use \cite{2016A&A...591A..50B}, which is specifically constructed to minimize CIB and SZ leakage, we verify that both this map and the SMICA map \citep{2020A&A...641A...4P} show no detectable LSS signals when cross-correlated with our reference galaxies on the relevant scales, indicating no risk of over-subtracting the signals of interest.

In addition to the data used in \citetalias{2020ApJ...902...56C}, we add three sky intensity maps from {\it Herschel}'s SPIRE instrument, the Spectral and Photometric Imaging Receiver at 250, 350, and 500~$\mu$m (1200, 857, and 600~GHz, respectively). We merge data taken as part of the Herschel Astrophysical Terahertz Large Area Survey \citep[H-ATLAS;][]{2016MNRAS.462.3146V,2017ApJS..233...26S}, and the Herschel Multi-tiered Extragalactic Survey \citep[HerMES;][]{2012MNRAS.424.1614O} using maps provided by the Centre d'Analyse de Donn\'ees Etendues (CADE)\footnote{\url{http://cade.irap.omp.eu}}. These maps are based on {\it Herschel}'s level 2.5 images, with HEALPix mosaics made using the resampling algorithm from \cite{2012A&A...543A.103P}. Unlike many other {\it Herschel} products optimized for source detection through aggressive local background subtraction, these maps preserve large-scale fluctuations of both the CIB and Milky Way foreground without a suppressed transfer function. Since we do not use small-scale information in the highly non-linear regime, we reduce the resolution from $N_{\rm side}=16384$ to 2048, matching that of the {\it Planck} and {\it IRAS} maps, which under-samples the {\it Herschel} beams. The effective beam FWHM is thus $1'.24$, set by the approximated Gaussian pixel window function. The sky coverage is not contiguous but spread across multiple fields totaling $1124\ \rm deg^2$. Among this area, $789\ \rm deg^2$ overlaps with the SDSS footprint, where we have reference sources for cross-correlations. We retain a subset of HerMES fields targeting lensing clusters, thus inherently overdense at the cluster redshifts. However, their total area is only $\sim 1$~deg$^2$, making any potential bias negligible.

We apply a common mask for bright point sources across all 11 maps, masking those detected in at least one {\it Planck} HFI channel \citep{2016A&A...594A..26P} or brighter than 1 Jy at 90~$\mu$m in the AKARI mission bright source catalog Version 2 \citep{2018cwla.conf..227Y}. We test the impact of varying the source masking thresholds on the CIB results in this paper and find them largely unchanged.  Nonetheless, masking bright sources reduces the frequency covariance in our CIB measurements.

\subsection{Foreground Mitigation}
\label{sec:foreground}

The Milky Way foreground from thermal emission of dust similar to the CIB introduces strong fluctuations on large scales. To mitigate, we mask $60\%$ of the sky with the highest dust column densities using the Milky Way dust map from \cite{2023ApJ...958..118C}, hereafter the CSFD map, smoothed over $2$~degrees. This reduces the usable sky area of {\it Planck} and {\it IRAS} accordingly but preserves most of the {\it Herschel} area, which has no low-latitude coverage to start with.

To further suppress foreground-induced noise and mitigate floating zero points in cross-correlations, \citetalias{2020ApJ...902...56C} used an HI map from \cite{2017ApJ...846...38L} for template-based cleaning. One might consider using far-infrared-based foreground templates, such as the \cite{1998ApJ...500..525S} dust map, for a more direct foreground subtraction. However, this would also remove the CIB, as these ``Galactic'' far-infrared products generally preserve extragalactic contamination \citep{2019ApJ...870..120C}. This issue is now resolved with the recently available CSFD Milky Way dust map, which is tomographically constructed to be CIB-free. We therefore use CSFD as a high-fidelity foreground template and regress out a linearly correlated component in each of our 11 far-infrared maps. This choice leads to smaller foreground residuals compared to cleaning using HI while fully preserving CIB fluctuations.

To further suppress the impact of residual foreground fluctuations, we downweight regions with high Galactic dust columns in our CIB-reference cross-correlations. Following Appendix D.2.2 and Figure~19 in \cite{2023ApJ...958..118C}, we test different powers of the $E(B-V)$ field in CSFD as the foreground suppressing weights for each pixel and find that choosing $E(B-V)^{-1.2}$ leads to a maximized S/N for CIB detection over $0<z<4$ in {\it Herschel} 250~$\rm \mu m$, a key band near the thermal CIB peak. For consistency, we adopt this weighting for all 11 bands. Note that since CSFD's $E(B-V)$ field traces only the foreground but not the CIB, this procedure enhances S/N purely by reducing noise without biasing or altering our cross-correlation amplitudes, as is indeed found in our test.

\begin{figure*}[t!]
    \begin{center}
    \includegraphics[width=0.985\textwidth]{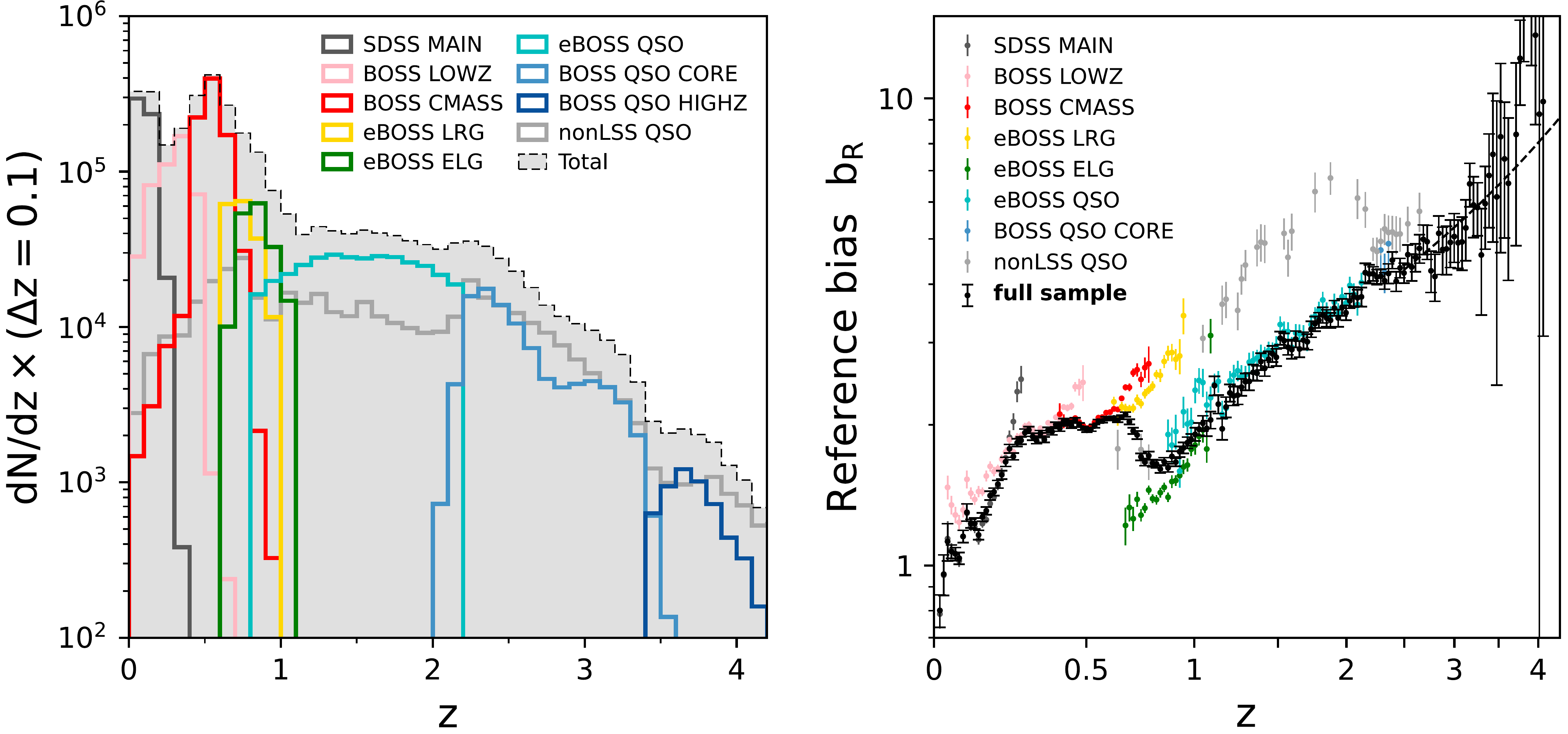}
    \end{center}
    \caption{Redshift distribution $dN/dz$ (left) and bias factor $b_{\rm R}$ (right) for the reference SDSS galaxies and QSOs. Their precise spectroscopic redshifts make them high-fidelity 3D tracers of LSS, enabling our CIB deprojection via cross-correlations. The $b_{\rm R}$ shown here is measured over scales of 0.5--10~Mpc~$h^{-1}$ across the full SDSS footprint. In practice, our CIB tomography uses multiple versions of the full-sample $b_{\rm R}(z)$, calculated based on the angular scales and footprint (with spatial weighting) applied for each frequency band. The dotted line shows a smooth function measured in \cite{2013ApJ...778...98S}, which we use at $z>2.5$ where our measurements become noisy. These $b_{\rm R}(z)$ are divided out from the raw galaxy–CIB cross-correlation amplitudes to isolate the tomographic CIB-only signal that will be shown in Figure~\ref{fig:dIdzb}.}
    \label{fig:sdss_ref}
\end{figure*}

\subsection{Cross-Correlation References}

As summarized in Figure~\ref{fig:sdss_ref}, we compile a set of spectroscopic reference galaxies and QSOs whose redshift information will be propagated into the CIB via angular cross-correlations. Following \cite{2023ApJ...958..118C}, we use extensively tested LSS data and random catalogs from SDSS, as well as the original and extended Baryon Oscillation Spectroscopic Surveys (BOSS, eBOSS). These include the SDSS MAIN galaxy sample \citep{2002AJ....124.1810S, 2005AJ....129.2562B, 2015MNRAS.449..835R}, BOSS LOWZ and CMASS luminous red galaxies \citep[LRGs;][]{2016MNRAS.455.1553R}, eBOSS LRGs, emission-line galaxies, and QSOs \citep{2020MNRAS.498.2354R,2021MNRAS.500.3254R}, as well as the BOSS CORE QSO sample \citep{2012MNRAS.424..933W,2015MNRAS.453.2779E}. To improve statistical power at high redshifts, we add two additional QSO samples: a non-LSS QSO catalog from SDSS Data Release 7 \citep{2010AJ....139.2360S} and high-redshift QSOs in BOSS that are not part of the CORE sample \citep{2017A&A...597A..79P}. For these two non-LSS QSO samples, for which random catalogs are not provided by the SDSS team, we construct additional random catalogs that account for the survey footprint while excluding veto areas affected by SDSS imaging artifacts and bright stars.

We combine these spectroscopic galaxies and QSOs into a single reference sample and merge the corresponding random catalogs, trimming them to maintain the same random-to-data ratio. The final reference sample includes $3,074,940$ spectroscopic redshifts up to $z\sim4.2$, spanning a redshift-dependent footprint within $\rm 10,440\ deg^2$ (approximately $1\pi$ steradian). Figure~\ref{fig:sdss_ref} (left panel) shows that the number of sources per redshift interval $\Delta z = 0.1$ is 1--$4\times10^5$ at $z<0.9$, about 10 times lower at $0.9 < z < 2.5$, and drops to $10^3$ at $z=4$. The main difference from the reference sample in \citetalias{2020ApJ...902...56C} is the use of the completed eBOSS LSS catalogs rather than an earlier release. Additionally, the handling of complex sample-dependent footprints and small-scale masks is improved by using random catalogs as opposed to pixelized selection function maps employed in \citetalias{2020ApJ...902...56C}.

\section{Measurement}\label{sec:result}

\subsection{Clustering Redshift}

Here, we measure the bias-weighted redshift distributions of EBL photons in each of the 11 bands using the clustering redshift, or cross-correlation redshift method, following the formalism in \cite{2019ApJ...877..150C} and \cite{2013arXiv1303.4722M}.

First, we bin our reference objects by redshift into 160 bins spaced in $\log(1+z)$ over $0 < z < 4.2$. This allows us to access their density contrast at each redshift $z_i$:
\begin{eqnarray}
\delta_{\rm R}(\phi, z_i) = \frac{n_{\rm R}(\phi, z_i) - \langle n_{\rm R}(z_i) \rangle}{\langle n_{\rm R}(z_i) \rangle},
\label{eq:delta_R}
\end{eqnarray}
where $\phi$ is the angular vector on the 2D sky, and $n_{\rm R}$ and $\langle n_{\rm R} \rangle$ represent the local and mean reference surface number densities, respectively. 

For each 2D intensity map $I_{\nu}(\phi)$, where the CMB has been removed and the Galactic foreground cleaned to first order using the CSFD template, we compute the excess intensity:
\begin{eqnarray}
\Delta I_{\nu}(\phi) = I_{\nu}(\phi) - \langle I_{\nu} \rangle
\label{eq:delta_I}
\end{eqnarray}
where $\langle I_{\nu} \rangle$ is the large-scale mean intensity, which we use local values calculated with a Gaussian kernel of 4~degree FWHM to be insensitive to foregrounds on scales beyond. Unlike the thin-sliced $\delta_{\rm R}$, here $\Delta I_{\nu}(\phi)$ is projected over a wide redshift range and may still contain residual foreground on small scales. We also keep $\Delta I_{\nu}(\phi)$ intentionally unnormalized, i.e., not divided by $\langle I_{\nu} \rangle$, to preserve the intensity units. $\Delta I_{\nu}$ thus resembles an excess density field for photons instead of a dimensionless density contrast more commonly used in cosmology.

\begin{figure*}[t!]
    \begin{center}
        \includegraphics[width=0.98\textwidth]{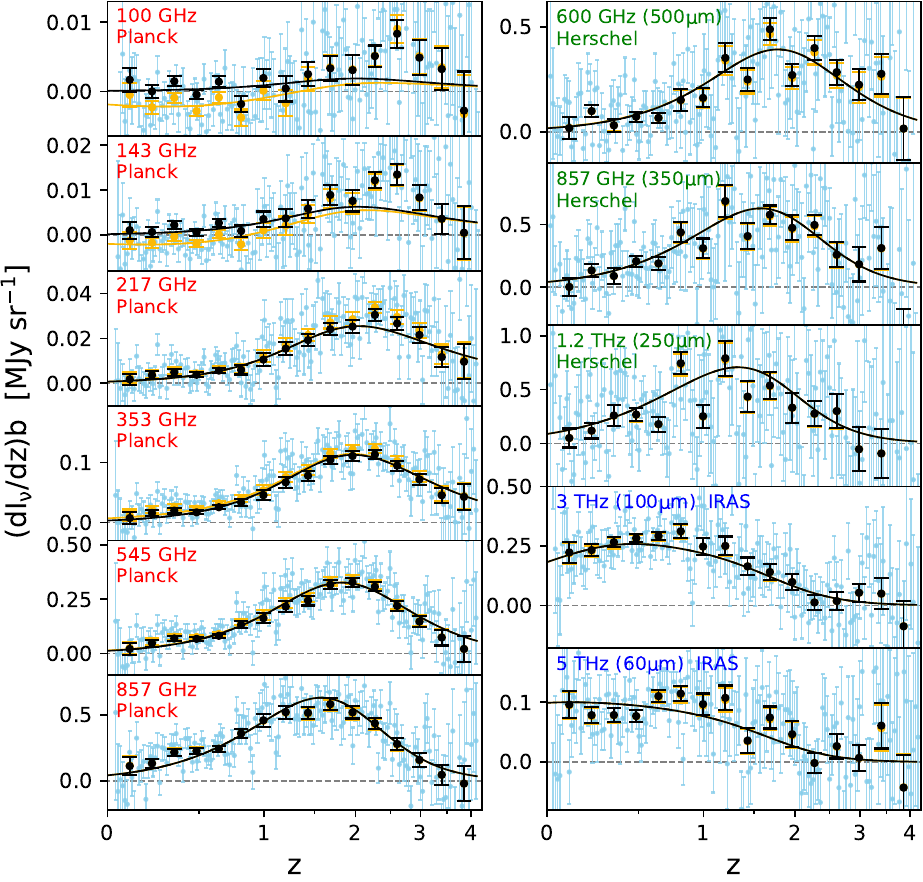}
    \end{center}
    \caption{Redshift-deprojected, bias-weighted mean CIB intensities $(dI_{\nu}/dz)\,b$ over $0<z<4$ for the 11 {\it Planck}, {\it Herschel}, and {\it IRAS} bands, corresponding to $\overline{w}_{\rm IR}/(b_{\rm R}\,\overline{w}_{\rm DM})$ in Equation~\ref{eq:clustering_z}. Light blue and yellow data points show raw measurements before and after redshift re-binning, both dominated by the CIB but also including the thermal SZ effect, more visible as decrements at 100 and 143~GHz at $z<1$. Black data points show the fiducial CIB result with two minor corrections: (1) SZ removal based on \citetalias{2020ApJ...902...56C} and (2) a color correction for deviations from the $\nu\,I_\nu = \mathrm{constant}$ in-band spectral shape assumed for the maps. Black (yellow) lines show the best-fit ensemble cosmic dust model without (with) SZ distortion. These redshift distributions can serve as baseline inputs for future studies using 2D far-infrared maps as 3D matter tracers.}
    \label{fig:dIdzb}
\end{figure*}

For each redshift bin $z_i$, we calculate $\overline{w}_{\rm IR}$, the angular cross-correlation function in real space between $\Delta I_{\nu}(\phi)$ and $\delta_{\rm R}(\phi, z_i)$. This series of cross-amplitudes is proportional to $(dI_{\nu}/dz)(z)$, the redshift distribution of the intensity field:
\begin{eqnarray}
\overline{w}_{\rm IR}(z_i) = 
\frac{dI_{\nu}}{dz}(z_i)\, b_{\rm I}(z_i) \, b_{\rm R}(z_i)\,\overline{w}_{\rm DM}(z_i)\;,
\label{eq:clustering_z}
\end{eqnarray}
where $b_{\rm I}$ and $b_{\rm R}$ are the effective linear bias factors for the intensity and reference sources, respectively, and $\overline{w}_{\rm DM}$ is the auto-correlation function of the underlying dark matter density field. In this equation, $\overline{w}_{\rm DM}$ can be computed given a cosmology, and $b_{\rm R}$ can be measured via calculating auto-correlations for the references. Correcting $\overline{w}_{\rm IR}(z)$ with these factors allows us to probe $(dI_{\nu}/dz)~b_{\rm I}(z)$, the bias-weighted redshift differential intensity. This tomographic property is intrinsic to the intensity field and is independent of the choice of reference sources after $b_{\rm R}$ de-biasing \citep[see Figure~12 in][]{2019ApJ...870..120C}.

The overline in $\overline{w}_{\rm IR}$ and $\overline{w}_{\rm DM}$ indicates that the correlation functions have been integrated over a predefined, quasi-linear angular scale $\theta$, following
\begin{eqnarray}
\overline{w} = \int_{\theta_{\rm min}}^{\theta_{\rm max}} W(\theta)\,w(\theta)\, d\theta\;,
\label{eq:wbar}
\end{eqnarray}
where $w(\theta) = w_{\rm IR}(\theta)$ or $w_{\rm DM}(\theta)$ is the full angular correlation function, and $W(\theta) \propto \theta^{-0.8}$ is a weight function following \cite{2013arXiv1303.4722M}. This weighting improves the S/N of $\overline{w}_{\rm IR}$ but does not introduce bias, as it is applied equally to $\overline{w}_{\rm DM}$ and cancels in Equation~\ref{eq:clustering_z}. The S/N is maximized when $W(\theta) \propto w(\theta)$; moderate deviations from this form are acceptable and result in only a slight reduction in S/N, while keeping the result largely unchanged. 

The matter correlation function $w_{\rm DM}$ is computed in each redshift bin using Equation (10) in \cite{2019ApJ...877..150C}, with a nonlinear Halofit \citep{2003MNRAS.341.1311S} power spectrum generated using the {\sf CLASS} code \citep{2011JCAP...07..034B, 2011JCAP...09..032L} under our assumed cosmology. As a post-processing step in this calculation, we also include a small suppression of $w_{\rm DM}$ to account for the use of 4-degree local $\langle I_{\nu} \rangle$ in Equation~\ref{eq:delta_I} as opposed to the cosmic mean. 

The cross-term $w_{\rm IR}$ is measured in the data as
\begin{eqnarray}
w_{\rm IR}(\theta, z_i)\  &=&\  \frac{1}{\langle Q(\phi) \rangle}\langle Q(\phi)\, \delta_{\rm R}(\phi, z_i) \cdot \Delta I_{\nu}(\phi+\theta) \rangle, 
\label{eq:weighted_cross}
\end{eqnarray}
where $Q$ is the foreground-suppressing weight map introduced in Section~\ref{sec:foreground} given by $Q(\phi) = E(B-V)(\phi)^{-1.2}$ from CSFD, which reduces the noise without altering the CIB signal. The dot product $\delta_{\rm R} \cdot \Delta I_{\nu}$ is key in deprojecting the CIB tomographically in redshift, where $\Delta I_{\nu}$ could be viewed as a composite vector and $\delta_{\rm R}$ the basis at thin redshift slides. The deprojection method is thus a simple vector decomposition. Additionally, any residual foregrounds not fully removed by template cleaning remain orthogonal to $\delta_{\rm R}$, contributing only noise without biasing $w_{\rm IR}$. 

To account for the selection function and mask of the reference sources, Equation~\ref{eq:weighted_cross} is evaluated twice: once using the data and once using the reference random catalog. The final $w_{\rm IR}$ is taken as the difference between these two measurements. This approach mirrors the \cite{1983ApJ...267..465D} estimator for galaxy two-point correlation functions, where a random catalog is used on one side of the cross-correlation.

For the integral in Equation~\ref{eq:wbar}, the lower bound ${\theta_{\rm min}}$ must be sufficiently large relative to the beam or pixel size, whichever is greater, and the physical size of dark matter halos to prevent contamination from the ``1-halo term'' clustering, which would break the linearity in Equation~\ref{eq:clustering_z}. However, accessing small scales is desirable for stronger signals. Given these considerations, we set a frequency- and redshift-dependent $\theta_{\rm min}$ corresponding to fixed physical scales of 2.5~Mpc$\,h^{-1}$ for 100~GHz, 2~Mpc$\,h^{-1}$ for 143~GHz, 1.5~Mpc$\,h^{-1}$ for {\it Planck} 217–857~GHz, 1~Mpc$\,h^{-1}$ for {\it IRAS} bands, and 0.5~Mpc$\,h^{-1}$ for {\it Herschel} bands. The upper bound, $\theta_{\rm max}$, is uniformly set to 10~Mpc$\,h^{-1}$ physical across all bands to mitigate wide-angle systematics. 

To further justify these scale cuts, we use the projected angular correlation function of detected {\it Herschel} sources from \cite{2010A&A...518L..22C} to estimate the 1-halo contamination near the peak redshift and frequency of the CIB. Given our $\theta^{-0.8}$ weighting, the 1-halo contribution is $\sim4\%$ for the most aggressive 0.5–10~Mpc$\,h^{-1}$ configuration used for {\it Herschel} bands, and drops to $\sim1\%$ for the less aggressive 1.5–10~Mpc$\,h^{-1}$ used for {\it Planck} 217–857~GHz, which provides most of the constraining power in this work thanks to {\it Planck}'s full-sky coverage. Since detected sources are the bright part of the CIB, these 1-halo estimates are conservative and can be viewed as upper limits. The linearity assumption in Equation~\ref{eq:clustering_z} over these scales has been tested in \cite{2022MNRAS.510.1223G} and \citetalias{2020ApJ...902...56C} (Appendix~A). However, since our $\theta_{\rm min}$'s are already in the non-linear regime, the proportionality in Equation~\ref{eq:clustering_z} holds only if a non-linear matter power spectrum is employed to compute $\overline{w}_{\rm DM}$, which is what we use (Halofit) in this work.

To determine the reference bias $b_{\rm R}$, we measure $\overline{w}_{\rm RR}$, the autocorrelation of the reference sample in each redshift bin $z_i$ and solve $\overline{w}_{\rm RR} = b_{\rm R}^2\,\overline{w}_{\rm DM}$ for each subset of our reference sample and for the full set as a whole. The results are shown in the right panel of Figure~\ref{fig:sdss_ref}. We use only the full-sample estimates in subsequent analyses, representing an effective bias weighted by the redshift-dependent abundances of different galaxy and QSO types. In addition to the overall increase with redshift, a prominent feature is the plateau of relatively high $b_{\rm R}$ at $0.25 < z < 0.7$ due to LRGs. Since the exact sample mixture is footprint-dependent and the combined bias is weakly scale-dependent, we repeat the $b_{\rm R}(z)$ measurements separately for each far-infrared band in the redshift tomography. The configuration matches the mission-dependent footprint, band-dependent scale cuts, and includes the spatial weighting adopted for foreground suppression. The statistical uncertainty in $b_{\rm R}$ is at the level of 1\% in our fiducial redshift binning (introduced later), which is coarser than that in Figure~\ref{fig:sdss_ref}, so we neglect the $b_{\rm R}$ errors in what follows.

After correcting for $b_{\rm R}$ and $\overline{w}_{\rm DM}$ in Equation~\ref{eq:clustering_z}, we obtain the tomographic EBL intensity $(dI_{\nu}/dz)\,b$ for each band (hereafter, we use $b$ as a shorthand for $b_{\rm I}$ for simplicity). We show the results in Figure~\ref{fig:dIdzb} in light blue data points with 160 densely sliced redshift bins. The CIB is detected across all bands at most redshifts, with especially high S/N in high-frequency bands in {\it Planck} and {\it IRAS} 100~$\mu$m. Error bars are derived empirically by resampling the sky 500 times using a block-bootstrap procedure, following \cite{2023ApJ...958..118C}. To reduce the size of the data vector, we further re-bin the original measurements, with inverse variance weighting, into 16 redshift bins within $0.05 < z < 4.1$ evenly spaced in $\log(1+z)$. The rebinned results are shown as yellow data points.

\begin{figure}[t!]
    \begin{center}
         \includegraphics[width=0.465\textwidth]{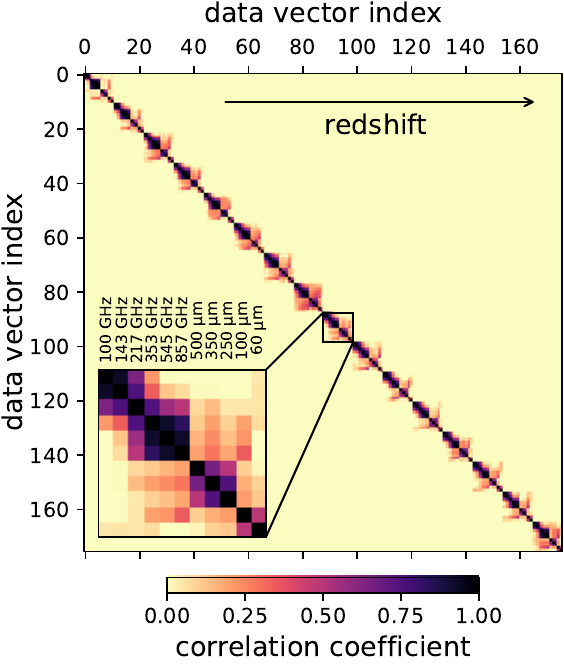}
    \end{center}
    \caption{Covariance matrix for our tomographic CIB measurements of 11 bands $\times$ 16 redshift bins, shown in a normalized, dimensionless form. The same correlation coefficients apply to both the $(dI_{\nu}/dz)\,b$ data vector in Figure~\ref{fig:dIdzb} and $\epsilon_{\nu}\,b$ in Figure~\ref{fig:rest_spectrum}.}
    \label{fig:covariance}
\end{figure}

We estimate the covariance matrix for the rebinned data vector of 11 bands times 16 redshifts using our 500 bootstrap samples and show the normalized correlation coefficient matrix in Figure~\ref{fig:covariance}. We find strong covariances in $(dI_{\nu}/dz)\,b$ across nearby frequencies at fixed redshift, likely driven by residual Galactic dust structures, with additional contribution from the intrinsic frequency coherence of the CIB. The three {\it Herschel} bands exhibit particularly strong internal covariances due to their smaller footprint, which enhances correlated foreground residuals and cosmic variance. These bands are only $\sim$30\% correlated with {\it Planck} at similar frequencies, suggesting that chance correlations between foreground residuals and SDSS references contribute significantly and largely de-correlate across different footprints. Not shown here, but we find a similar level of de-correlation within the same {\it Planck} map when comparing regions inside and outside a {\it Herschel}-sized sky. For elements at different redshifts $z_i \neq z_j$, i.e., outside the 11-band diagonal, we expect no correlation, as the redshift bin widths are much larger than the typical LSS correlation length. The measured covariances for these elements are consistent with zero, though not exactly so, which is expected since the measured ``errors of errors'' are inherently noisier. For this reason, we set all off-redshift covariances to zero. One could consider the final covariance matrix as a hybrid one with empirical and theoretical components, where the theory part is simply zero for $z_i \neq z_j$ elements. 

\subsection{SZ Removal and Color Correction}
Our raw $(dI_{\nu}/dz)\,b$ measurements probe the full EBL, capturing all emission mechanisms combined. This includes both the CIB and the thermal SZ effect, i.e., the Compton $y$ spectral distortion off the CMB by hot gas in the LSS:
\begin{equation}
\frac{dI_{\nu}}{dz}\,b\,(\nu, z) = \left[\frac{dI_{\nu}}{dz}\,b\,(\nu, z)\right]_{\rm SZ} + \left[\frac{dI_{\nu}}{dz}\,b\,(\nu, z)\right]_{\rm CIB}.
\label{eq:tot_is_cib_plus_sz}
\end{equation}
The SZ effect is weak compared to the CIB and is noticeable only at low frequencies, as seen in the decrements, i.e., negative yellow data points in Figure~\ref{fig:dIdzb} at 100~GHz at $z<1$. The bias-weighted SZ distortion is given by:
\begin{equation}
\left[\frac{dI_{\nu}}{dz}\,b\,(\nu, z)\right]_{\rm SZ} =  \frac{dy}{dz}(z)\,b_y(z)\, g(x)\, I_{\nu 0}\,,
\label{eq:y_distortion}
\end{equation}
where $y = (\sigma_{\rm T}/m_{\rm e}\, c^2) \int d\chi\, (1+z)^{-1}\, P_{\rm e}(\chi)$ defines the Compton $y$ parameter \citep{2002ARA&A..40..643C}, with $\sigma_{\rm T}$, $m_{\rm e}$, $c$, $k_{\rm B}$, $\chi$, and $P_{\rm e}$ denoting the Thomson cross section, electron mass, speed of light, Boltzmann constant, comoving radial distance, and the electron pressure of hot gas, respectively. The bias factor $b_y$ is frequency-independent and is calculated in \citetalias{2020ApJ...902...56C} using a halo model, more specifically, as a $y$-weighted linear bias of halos hosting the hot gas. In Equation~\ref{eq:y_distortion}, $I_{\nu 0} = 2\,(k_{\rm B}\,\textrm{T}_{\rm CMB})^3/(hc)^2$ is the characteristic amplitude per unit of $y$, and
\begin{equation}
g(x) = \frac{x^4\, e^x}{(e^x-1)^2}\, \Bigg( x\, \frac{e^x+1}{e^x-1}-4 \Bigg)
\label{eq:g_x}
\end{equation}
is the frequency dependence with $x \equiv h\,\nu/(k_{\rm B}\,\textrm{T}_{\rm CMB})$, assuming a CMB temperature of $\textrm{T}_{\rm CMB} = 2.725$~K. To isolate the CIB spectrum in this work, we remove the SZ term, redshift by redshift, using the best fit $(dy/dz)\,b_y(z)$ constrained by the tomographic measurements in \citetalias{2020ApJ...902...56C}.

\tabcolsep=0.162cm % This reduce table width to fit text width in this case
\begin{table*}[t!]
\begin{center}
%\centering
\caption{\label{table:per_band_stats} \normalsize   CIB Monopole from Integrating Redshift Tomographic Intensities [MJy sr$^{-1}$]}
\renewcommand{\arraystretch}{1.15}
%\hspace{-1.6cm}
\begin{tabular}{c|cccca|c|a}
\hline
\rowcolor{DarkerGray}[\tabcolsep]
Band & \multicolumn{5}{c|}{CIB} & SZ & CIB$\ + \ $SZ
\\ 
\hline
 & S/N$^{\ \mathtt{a}}$ & $I_{\nu}\langle b \rangle^{\ \mathtt{b}}$ & $I_{\nu}^{\ \mathtt{c}}$ & $I_{\nu}^{\ \mathtt{d}}$ & $I_{\nu}^{\ \mathtt{e}}$ &  $I_{\nu}^{\ \mathtt{f}}$ & $I_{\nu}^{\ \mathtt{g}}$
\\ 
 &   & direct 0$<$$z$$<$4 & direct 0$<$$z$$<$4 & direct total&  global fit & Chiang+20 & global fit 
\\
\hline
{\it Planck} 100~GHz & 7.3  &  $0.010\pm{0.003}$ & $0.004\pm{0.001}$ & $0.005\pm{0.001}$ & $0.0023\pm{0.0003}$ & $-0.0012\pm{0.0002}$ & $0.0011\pm{0.0003}$ \\
{\it Planck} 143~GHz & 12.4  & $0.023\pm{0.003}$ & $0.009\pm{0.001}$ & $0.011\pm{0.001}$ & $0.0076\pm{0.0006}$ & $-0.0013\pm{0.0002}$ & $0.0063\pm{0.0007}$ \\
{\it Planck} 217~GHz & 23.8  & $0.064\pm{0.004}$ & $0.026\pm{0.001}$ & $0.031\pm{0.002}$ & $0.0297\pm{0.0018}$ & $...$ & $0.0297\pm{0.0018}$ \\
{\it Planck} 353~GHz & 34.0  & $0.256\pm{0.012}$ & $0.106\pm{0.004}$ & $0.118\pm{0.005}$ & $0.122\pm{0.005}$ & $0.0022\pm{0.0004}$ & $0.124\pm{0.005}$ \\
{\it Planck} 545~GHz & 36.1  & $0.658\pm{0.034}$ & $0.294\pm{0.012}$ & $0.307\pm{0.013}$ & $0.326\pm{0.010}$ & $0.0011\pm{0.0002}$ & $0.327\pm{0.010}$ \\
{\it Planck} 857~GHz & 34.2  & $1.150\pm{0.073}$ & $0.577\pm{0.027}$ & $0.584\pm{0.027}$ & $0.599\pm{0.021}$ & $...$ & $0.599\pm{0.021}$ \\
 \hline
 {\it Herschel} 500~$\rm \mu m$ & 17.3  & $0.886\pm{0.087}$ & $0.378\pm{0.029}$ & $0.390\pm{0.030}$ & $0.385\pm{0.012}$ & $0.0007\pm{0.0001}$ & $0.386\pm{0.012}$ \\
 {\it Herschel} 350~$\rm \mu m$ & 16.4  & $1.078\pm{0.147}$ & $0.535\pm{0.050}$ & $0.542\pm{0.051}$ & $0.599\pm{0.021}$ & $...$ & $0.599\pm{0.021}$ \\
 {\it Herschel} 250~$\rm \mu m$ & 13.0  & $0.610\pm{0.214}$ & $0.443\pm{0.071}$ & $0.445\pm{0.071}$ & $0.681\pm{0.029}$ & $...$ & $0.681\pm{0.029}$ \\
  \hline
 {\it IRAS} 100~$\rm \mu m$ & 32.7  &   $0.473\pm{0.058}$ & $0.312\pm{0.021}$ & $0.313\pm{0.021}$ & $0.286\pm{0.019}$ & $...$ & $0.286\pm{0.019}$ \\
 {\it IRAS} 60~$\rm \mu m$ & 21.2  & $0.191\pm{0.032}$ & $0.119\pm{0.011}$ & $0.119\pm{0.011}$ & $0.110\pm{0.012}$ & $...$ & $0.110\pm{0.012}$ \\
\hline
\multicolumn{8}{l}{}
\vspace{-0.3cm}\\
\multicolumn{8}{l}{$^{\mathtt{a}}$ Signal-to-noise ratio for CIB tomography $(dI_{\nu}/dz)\, b$ per band.} \\
\multicolumn{8}{l}{$^{\mathtt{b}}$ Direct detection of bias-weighted CIB monopole up to $z=4$, which is $\int_{z=0}^{z=4} (dI_{\nu}/dz)\,b\, dz$, integral of black data points in Figure~\ref{fig:dIdzb}.}\\
\multicolumn{8}{l}{$^{\mathtt{c}}$ Direct detection of CIB monopole up to $z=4$, or $\int_{z=0}^{z=4} (dI_{\nu}/dz)\, dz$. The bias is corrected (uncertainty added) using $b(z)$ in Figure~\ref{fig:bias}.} \\
\multicolumn{8}{l}{$^{\mathtt{d}}$ Total CIB monopole $\int_{z=0}^{z=\infty} (dI_{\nu}/dz)\, dz$ by applying a small correction factor $1/f_{\rm det}$ to $I_{\nu}^{\ \mathtt{c}}$, with $f_{\rm det}$ from the top panel in Figure~\ref{fig:monopole}.}\\
\multicolumn{8}{l}{$^{\mathtt{e}}$ Ensemble dust  model in Section~\ref{sec:model} fit to $(dI_{\nu}/dz)\,b$'s plus external monopoles up to 545~GHz; same as the total (red line) in Figure~\ref{fig:monopole}. }\\
\multicolumn{8}{l}{$^{\mathtt{f}}$ Thermal SZ from hot gas in LSS scattering CMB. Total $y = (1.22 \pm 0.20) \times 10^{-6}$ \citepalias{2020ApJ...902...56C}, causing small distortions compared to CIB.}\\
\multicolumn{8}{l}{$^{\mathtt{g}}$ Total EBL = CIB ($I_{\nu}^{\ \mathtt{e}}$) + SZ ($I_{\nu}^{\ \mathtt{f}}$), same as the cyan line in Figure~\ref{fig:monopole}. \cite{2019ApJ...877...40O} 100--545 GHz monopoles are used in the fitting.

}\\
\end{tabular}
\end{center}
\end{table*}

We apply a ``color correction'' (CC) to the CIB alongside SZ removal. The goal of CC is to report $(dI_{\nu}/dz)\,b$ at the precise central frequency of each band while incorporating the best knowledge of the in-band spectral shape. Our far-infrared maps, and thus the raw $(dI_{\nu}/dz)\,b$ values shown as yellow data points in Figure~\ref{fig:dIdzb}, follow the {\it IRAS} convention, which assumes $\nu I_\nu = \mathrm{constant}$ in-band. After performing multi-band tomography, we can refine this assumption in a frequency- and redshift-dependent manner.  

For each redshift bin, we fit the 11 $(dI_{\nu}/dz)\,b$ data points with a smooth CIB SED in the rest frame plus the observer-frame SZ distortion with amplitude $dy/dz$ fixed to that in \citetalias{2020ApJ...902...56C}. During this fitting, we forward-model the CC effect from high-resolution SED to broadband photometry under the {\it IRAS} convention, following the procedure in \cite{2014A&A...571A...9P} and accounting for precise filter transmission curves.\footnote{For {\it Herschel} SPIRE bands, we use the response functions for extended source calibration, as described in \url{http://herschel.esac.esa.int/Docs/SPIRE/pdf/spire_om_v24.pdf}.} The choice of the CIB SED fitting function at this stage is somewhat arbitrary and need not be physical, as long as it provides good fits to the already well-sampled data.\footnote{In fact, a model that overfits the data or an interpolation works just as fine.} Here, we use a modified blackbody with a lognormal power-law temperature distribution, which will be formally introduced in Section~\ref{sec:model_warm_cold}, but we have verified that the derived CC factors remain consistent when using alternative SED models with similar flexibility. We then divide out these CC factors, point by point, for $(dI_{\nu}/dz)\,b$ in Figure~\ref{fig:dIdzb}.

We present the SZ-removed, color-corrected $(dI_{\nu}/dz)\,b$ measurements as black data points in Figure~\ref{fig:dIdzb}. Compared to the uncorrected values (yellow points), the SZ effect is noticeable only at low frequencies at $z\lesssim1$, while the remaining difference is due to CC. Although both corrections are small in amplitude, they are necessary for achieving accurate measurements and a robust interpretation of the CIB, given our overall high S/N.

\subsection{Results---Tomographic CIB Spectrum}\label{sec:result_CIB_spectrum}

As shown in Figure~\ref{fig:dIdzb}, the CIB is detected across all frequency bands, and our redshift coverage up to $z=4$ is sufficient to capture the rise and fall of the CIB over cosmic time. We list the per-band significance in Table~\ref{table:per_band_stats}, where the S/N exceeds 30 for {\it Planck} 353, 545, 857~GHz and {\it IRAS} 100~$\mu$m. The combined significance across all 11 bands is 58, calculated as $\rm{S/N} = ($$\bm{d}^T\,\bm{C}^{-1}\, \bm{d})^{1/2}$, where $\bm{d}$ is the data vector of length $176 = 11$ bands $\times$ 16 redshifts, and $\bm{C}$ is the covariance matrix from Figure~\ref{fig:covariance}. In Figure~\ref{fig:dIdzb}, a strong negative $K$-correction is evident at low frequencies, where the CIB peaks at higher redshifts. These $(dI_{\nu}/dz)\,b$ results can benefit a wide range of future cosmological applications using CIB as a matter tracer, such as CIB-CMB lensing and CIB-galaxy clustering, where band-dependent redshift kernels are required for interpretation. Additionally, integrating $(dI_{\nu}/dz)\,b$ over $0<z<4$ provides maximally empirical, foreground- and background-free measurements of the bias-weighted CIB monopole in each band, which we list as $I_{\nu}\langle b \rangle$ in Table~\ref{table:per_band_stats}.

\begin{figure}[t!]
    \begin{center}
         \includegraphics[width=0.465\textwidth]{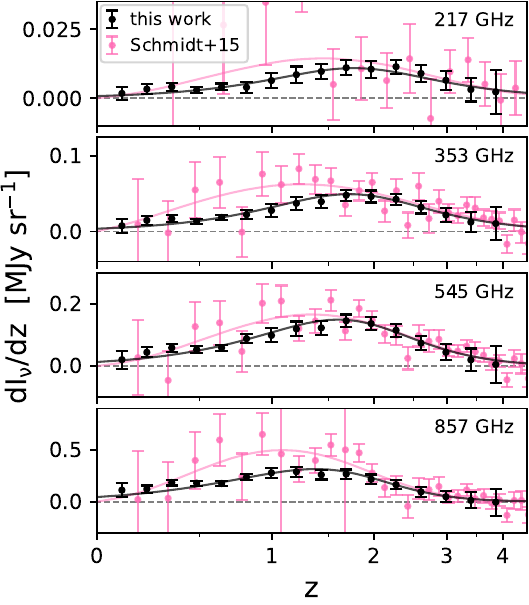}
    \end{center}
    \caption{Comparison of our bias-corrected $dI_{\nu}/dz$ (black) with that from \cite{2015MNRAS.446.2696S} (pink) for the 4 {\it Planck} bands measured in the latter. Our results, taken from Figure~\ref{fig:dIdzb} with bias removed using Figure~\ref{fig:bias}, have much higher S/N. \cite{2015MNRAS.446.2696S} likely overestimate CIB amplitudes at $z<2$ due to the inclusion of small-scale clustering within and around the {\it Planck} beams.}
    \label{fig:compare_Schmidt15}
\end{figure}

\begin{figure*}[t!]
    \begin{center}
         \includegraphics[width=0.85\textwidth]{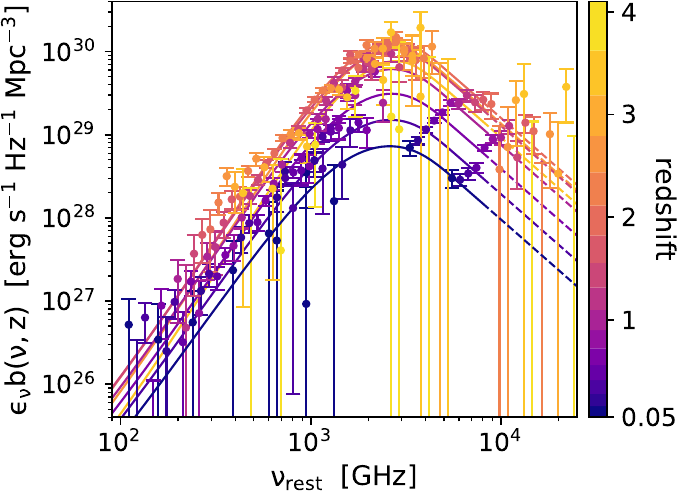}
    \end{center}
    \caption{Tomographic CIB spectrum in the rest frame, expressed as the bias-weighted CIB emissivity $\epsilon_{\nu}\,b$ over $0<z<4$. Data points are derived from the observed-frame $(dI_{\nu}/dz)\,b$ in Figure~\ref{fig:dIdzb} (black data points) via a simple $z$-dependent unit conversion and $1+z$ frequency shifts, leading to dense spectral sampling. The overall amplitude increases from $z=0$ to $z=2$--3 before declining at $z>3$ (occurring at a higher redshift than the peak cosmic star formation since $b(z)$ is monotonically increasing). Solid lines, colored by redshift, show our best-fit ensemble cosmic dust SEDs from the same posterior as Figure~\ref{fig:dIdzb} but sliced along different dimensions. The model is flexible enough for the best fits to smoothly interpolate the data. The peak frequency, and thus dust temperature, increases slightly toward higher redshifts. This tomographic CIB spectrum forms the foundation for a census of cosmic dust and star formation.}\
    \label{fig:rest_spectrum}
\end{figure*}

We evaluate internal consistency by comparing our measurements in {\it Herschel} 350~$\mu$m and {\it Planck} 857~GHz, which have nearly identical effective frequencies but differ in sky coverage and survey depth. The beam sizes also differ significantly, resulting in different angular scale cuts $\theta_{\rm min}$ for cross-correlations. Encouragingly, the measured amplitudes in these two bands are consistent in both the tomographic $(dI_{\nu}/dz)\,b$ evolution in Figure~\ref{fig:dIdzb} and the redshift-integrated monopole in Table~\ref{table:per_band_stats}, reinforcing our confidence in recovering unbiased CIB statistics despite strong foregrounds. Since the particular processing of {\it Herschel} data has been less explored in intensity mapping analyses, this agreement with an anchor frequency from {\it Planck} provides strong validation for the other two {\it Herschel} bands, particularly 250~$\mu$m, which bridges the gap between {\it Planck} and {\it IRAS} in probing the peak CIB frequency.

We compare our results with the seminal work of \cite{2015MNRAS.446.2696S}, who applied a similar method to four {\it Planck} bands using SDSS DR7 QSOs as cross-correlation references. \cite{2015MNRAS.446.2696S} derived $dI_{\nu}/dz$ from tomographic CIB-QSO clustering, correcting for CIB bias by incorporating per-channel auto power spectra. However, they assumed that the CIB and QSOs share the same bias factor, an approximation that will break down at high S/N. In Figure~\ref{fig:compare_Schmidt15}, we overlay their $dI_{\nu}/dz$ (pink)\footnote{We find inconsistencies within \cite{2015MNRAS.446.2696S} between their Figure~1, Table~2, and Table~B2; Figure~1 is used for this comparison.} with ours (black) for the four {\it Planck} bands in common. To correct for the CIB bias in our result, we use $b(z)$ posterior in Figure~\ref{fig:bias}, which combines direct monopole measurements with our redshift tomography. This will be detailed in Section~\ref{sec:model}, but the key point is the use of a more empirically constrained $b(z)$. The overall S/N in \cite{2015MNRAS.446.2696S} is lower, and their best-fit model overestimates the CIB amplitude at $z\sim1$ by a factor of two, a marginally significant bias given the same behavior across all four bands. This discrepancy may be related to their choice of an aggressive scale cut ($\theta_{\rm min} = 0.3$~Mpc, compared to 2.2~Mpc in our analysis), which extends well into the {\it Planck} beam. As a result, 1-halo term signals, i.e., dust emission from QSO hosts rather than the bulk of the CIB, likely leaked into their estimates, breaking the main linearity assumption in clustering redshift (Equation~\ref{eq:clustering_z}).

We can also compare our $(dI_{\nu}/dz)\,b$ values without bias correction to those in \citetalias{2020ApJ...902...56C} (Figure~2 therein) for the eight {\it Planck} and {\it IRAS} bands in common. We find the results to be consistent and highly correlated, as both works share a large fraction of the SDSS reference galaxies and use intensity maps based on the same product, albeit with different foreground mitigation schemes. The SZ distortion at the lowest frequencies is retained in \citetalias{2020ApJ...902...56C}, whereas in this work, we remove it to focus on the CIB. The amplitude of this SZ correction is small, as seen in the difference between the yellow and black data points in Figure~\ref{fig:dIdzb} for the 100~GHz band at the peak of the SZ decrement.  

All in all, with improved frequency coverage, foreground mitigation, clustering redshift analyses, and overall S/N, we consider our new results to supersede those in \cite{2015MNRAS.446.2696S} and \citetalias{2020ApJ...902...56C}.

For the CIB as an ensemble, we now combine the multi-band $(dI_{\nu}/dz)\,b$ in Figure~\ref{fig:dIdzb} to form a single, evolving spectrum or a bias-weighted emissivity $\epsilon_{\nu}\,b$ in Figure~\ref{fig:rest_spectrum}. This is simply a unit conversion from the observer to the rest frame using the cosmological radiative transfer equation:
\begin{equation}
\epsilon_{\nu}\,b\, (\nu_{\rm rest}, z) =  \frac{4\pi\, H(z)\, (1+z)}{c}\,\frac{dI_{\nu}}{dz}\,b\,(\nu_{\rm obs}, z)\,,
\label{eq:dIdz_to_emissivity}
\end{equation}
where $\nu_{\rm rest} = (1+z)\, \nu_{\rm obs}$, $c$ is the speed of light, and $H(z)$ is the Hubble parameter. Figure~\ref{fig:rest_spectrum} also shows that our redshift tomography provides enhanced sub-bandpass frequency information, as demonstrated in a ``spectral tagging'' analysis in \cite{2019ApJ...877..150C}. Each observed frequency band now spawns 16 sampling in rest frequencies shifted by a series of $1+z$ factors. One notable example is the high S/N track of the {\it IRAS} 100~$\mu$m band, which shifts from 3~THz at $z=0$ to the upper right in Figure~\ref{fig:rest_spectrum}. On the Rayleigh-Jeans side, adjacent {\it Planck} bands begin to overlap in the frequency-redshift space. As a result, we achieve a denser frequency sampling in our tomographic CIB spectrum than that of any single dusty galaxy SED reported in the literature.

In the CIB spectrum in Figure~\ref{fig:rest_spectrum}, the dynamic range of $\epsilon_{\nu}\,b$ spans more than four orders of magnitude ($10^{26}$--$10^{30}$ $\rm erg\ s^{-1}\ Hz^{-1}\ Mpc^{-3}$). The overall amplitude, and thus the total energy budget, increases from $z=0$ toward about $z=2$ before declining at $z > 3$. The spectral shape clearly indicates a thermal origin. We find a peak frequency at 2--3 THz, slightly higher at high redshifts, implying a temperature evolution. This data-driven tomographic CIB spectrum can serve as a valuable benchmark for future studies in precision cosmology and astrophysics. We thereby make the full data vector and covariance publicly available.\footnote{\label{fn:datarelease}\url{https://zenodo.org/records/16486649}}

\section{Astrophysical Information}\label{sec:interpretation}

We now extract astrophysical information from the tomographic CIB spectrum in Figure~\ref{fig:rest_spectrum}.

\subsection{Ensemble Cosmic Dust SED Fitting}\label{sec:model}

The far-infrared SEDs for both individual dusty galaxies and the CIB are routinely modeled as a single-temperature modified blackbody \citep[MBB; e.g.,][]{2002PhR...369..111B,2014PhR...541...45C,2020A&A...641A...4P}, parameterized by temperature $T$, opacity spectral index $\beta$, and a normalization. However, an MBB is known to be insufficient even for single-galaxy spectra when frequency sampling is sufficiently broad. Dust in the multi-phase ISM generally exhibits a temperature distribution, and even a single resolved sightline can contain multiple components \citep{2015ApJ...798...88M,2022arXiv221107667Z}. Beyond the ISM, the presence of significant dust in the circumgalactic (CGM) or intergalactic (IGM) medium remains uncertain. If such dust exists, its emission would contribute to the CIB.\footnote{Even for cold dust at $\textrm{T}_{\rm CMB}$, our method could still detect emission correlated with the LSS without being overshadowed by the primary CMB field.} In addition to variations on galactic and circumgalactic scales, diversity across galaxy populations leaves further imprints on the CIB.

These complexities necessitate a more realistic SED model. Indeed, we find that our CIB spectrum in Figure~\ref{fig:rest_spectrum} cannot be fitted by a single-component MBB, even when allowing redshift evolution for both $T$ and $\beta$. We thus generalize MBB model to describe a realistic ensemble of cosmic dust, requiring sufficient flexibility to capture the full distributions of $T$ and $\beta$ and their evolution. Since information on the host environments (e.g., ISM, CGM, or IGM) is inferred only through temperature, we adopt the classic temperature nomenclature \citep{2005astro.ph..3571L}: cold dust (10--30~K), warm dust (30--1000~K), and very cold dust ($<10$~K), while using these classifications somewhat loosely as the physical picture is continuous. The bulk of the CIB emissivity originates from the cold and warm phases, while very cold dust, though may be present in the ISM, contributes negligibly to the total mass \citep{1995ApJ...451..188R}. In what follows, we construct a fiducial warm-cold dust model to be as inclusive as possible for a census of dust in the universe. Some very cold dust could be missing only if the temperature distribution exhibits a bimodality, which we will discuss later in an extended model.

Since our data vector is the cosmic mean spectrum (Figure~\ref{fig:rest_spectrum}), with spatial anisotropy already consumed, there is no need to invoke a ``halo model.'' Instead, we simply treat the ensemble of cosmic dust as a single entity for SED fitting. Below we specify each element needed to fit our main measurements $\epsilon_{\nu}\,b$ in Figure~\ref{fig:rest_spectrum}.

\subsubsection{Effective Bias}\label{sec:model_bias}

Our tomographic CIB spectrum is weighted by the effective bias, which is, by definition, the overdensity ratio between the CIB and underlying matter. This bias factor can be interpreted as a light-weighted galaxy bias, providing insights into galaxy-halo connection for the CIB. To incorporate it into the inference, we assume the bias to be frequency-independent while allowing for a flexible redshift evolution
\begin{eqnarray}
b(z) = b_0 + b_1\,z + b_2\,z^2 \,.
\label{eq:b_of_z}
\end{eqnarray}
Among the three free parameters, $b_0$, $b_1$, and $b_2$, at least one effective degree of freedom (DOF) can be empirically constrained by incorporating the CIB monopole measured in the literature (specified later) into the data vector \citep[see Section~4.2.2 in][]{2019ApJ...877..150C}. The remaining two can be treated as nuisance parameters and marginalized over during our inference.

\subsubsection{Warm-Cold Dust}\label{sec:model_warm_cold}

We now model the ensemble dust SED. For a general opacity without enforcing an optically thin assumption, the CIB emissivity from the thermal continuum of warm-cold cosmic dust is given by:
\begin{eqnarray}
\epsilon_{\nu} (\nu)\, &\equiv& \,\frac{dL_{\nu}(\nu)}{dV} \nonumber\\ 
&=&\, \frac{4\pi\,dA_{\rm d}}{dV}\, (1-e^{- \langle \tau (\nu)\rangle}) \, \langle B_{\nu}(\nu)\rangle \nonumber\\ 
&=&\, \frac{4\pi\, \rho_{\rm{d}}}{\Sigma_{\rm d}}\, (1-e^{- \langle \tau (\nu)\rangle}) \, \langle B_{\nu}(\nu)\rangle\, ,
\label{eq:SED_1}
\end{eqnarray}
with redshift dependence allowed in all elements not explicitly shown for simplicity. In this expression, $L_{\nu}$ is the specific luminosity, $dV$ is the comoving volume element, and $dA_{\rm d}/dV$ represents the effective dust emission surface area per unit cosmic volume. $\langle \tau \rangle$ and $\langle B_{\nu} \rangle$ are the population-weighted optical depth and source function, respectively, which will be specified. The last line introduces a key quantity of interest—the comoving dust volume density, $\rho_{\rm d} = dM_{\rm d} / dV$, in $\rm M_{\odot}\ Mpc^{-3}$, where $dM_{\rm d}$ is the cosmic dust mass element. The dimensionless cosmic dust density parameter, $\Omega_{\rm dust}$, is traditionally defined as
\begin{eqnarray}
\Omega_{\rm dust} = \frac{\rho_{\rm d}}{\rho_{\rm crit}}\,,
\label{eq:omega_dust}
\end{eqnarray}
where $\rho_{\rm crit}$ is the present-day critical density. In Equation~\ref{eq:SED_1}, we also define an effective dust surface density, $\Sigma_{\rm d} = dM_{\rm d} / dA_{\rm d}$, in $\rm M_{\odot}\ kpc^{-2}$, which is needed to relax the commonly used optically thin assumption and allow for general opacity. We note that $\rho_{\rm d}$, and thus $\Omega_{\rm d}$, is a global quantity as the volume element in the denominator is the cosmic volume by definition. In contrast, $\Sigma_{\rm d}$ is a local quantity, as it is related to the effective size of individual galaxies or the typical dust emission area. 

The physical role of $\Sigma_{\rm d}$ lies in the effective optical depth:
\begin{eqnarray}
\langle \tau(\nu) \rangle 
\,&=&\, \kappa_0 \, \Sigma_{\rm d} \, \biggl \langle \left(\frac{\nu}{\nu_0}\right)^{\beta} \biggl \rangle \nonumber \\
\,&=&\, \kappa_0 \, \Sigma_{\rm d} \, \int \left(\frac{\nu}{\nu_0}\right)^{\beta} \, P(\beta)\, d\beta\,,
\label{eq:tau_effective}
\end{eqnarray}
where $\nu_0 = 353$~GHz ($\lambda_0 = 850$~$\mu$m) is an arbitrary reference frequency, and $\kappa_0$ is the opacity at $\nu_0$. We note that $\kappa_0$ is highly uncertain but could be viewed as a normalization factor or a dust mass-to-light ratio anchor to interpret the CIB amplitudes. For a more direct comparison with literature results, we adopt one of the most commonly used values, $\kappa_0 = 0.77~\rm cm^2\,g^{-1}$, at $850$~$\mu$m \citep{2000MNRAS.315..115D}. The spectral index $\beta$, as in the standard MBB model, characterizes the optical properties of dust grains. The angle brackets in Equation~\ref{eq:tau_effective} indicate that the frequency dependence of the optical depth is also an effective one, weighted by a distribution of $\beta$ for the ensemble dust, which we assume to follow a Gaussian distribution with mean $\langle \beta \rangle$ and spread $\sigma_{\beta}$:
\begin{eqnarray}
P(\beta)\,\propto\, e^{-\frac{1}{2}\left(\frac{\beta - \langle \beta \rangle}{\sigma_{\beta}}\right)^2}. 
\label{eq:P_beta}
\end{eqnarray}

It is informative to examine the limiting cases in the optically thin and thick regimes:
\begin{eqnarray}
  \epsilon_{\nu} =
  \begin{dcases}
  4\pi\, \rho_{\rm{d}}\, \kappa_0 \,\biggl \langle \left(\frac{\nu}{\nu_0}\right)^{\beta} \biggl \rangle \, \langle B_{\nu}\rangle\ & \text{     if $\langle\tau\rangle \ll 1$\,,} \\
  \frac{4\pi\, \rho_{\rm{d}}}{\Sigma_{\rm d}}\, \langle B_{\nu}\rangle\  & \text{     if $\langle\tau\rangle \gg 1\,,$}
  \end{dcases}
  \label{eq:optical_thin_thick}
\end{eqnarray} 
where the two can be equivalently discussed in wavelength at $\gg \lambda_{\rm thick}$ (thin) and $\ll \lambda_{\rm thick}$ (thick) regimes, with $\lambda_{\rm thick}$ denoting the characteristic wavelength at $\langle\tau\rangle=1$. In the optically thick regime or shorter wavelengths, the dust mass density $\rho_{\rm d}$, and thus the derived $\Omega_{\rm d}$, could be degenerate with $\Sigma_{\rm d}$. Fortunately, this degeneracy is broken at long wavelengths or low frequencies. Our choice to set $\Sigma_{\rm d}$ a free parameter thus allows $\lambda_{\rm thick}$ to be empirically constrained by the data. An approximate scaling follows:
\begin{equation}
\lambda_{\rm thick} \sim (\kappa_0\,\Sigma_{\rm d})^{1/\beta}\,\lambda_0\,,
\label{eq:lambda_thick}
\end{equation}
which works precisely only in the single-component MBB case but offers useful insight. We note that the absolute dust mass constraints still depend on the assumed reference opacity $\kappa_0$, with an approximate scaling of $\Omega_{\rm d} \propto \kappa_0^{-1}$. For comparisons with literature using different $\kappa_0$ values, one needs to rescale accordingly.  

We now specify the source function $\langle B_{\nu} \rangle$ in Equation~\ref{eq:SED_1}, which is weighted by a distribution of temperatures $P(T)$:
\begin{eqnarray}
\langle B_{\nu} (\nu)\rangle  = \int B_{\nu}(\nu, T)\,P(T)\, dT\,, 
\label{eq:B_effective}
\end{eqnarray}
where $B_{\nu}$ is the Planck function:
\begin{eqnarray}
B_{\nu}(\nu, T) = \frac{2 h \nu^3}{c^2} \frac{1}{e^{\frac{h \nu}{k_B T}} - 1}, 
\label{eq:Planck_fn}
\end{eqnarray}
with $h$, $c$, and $k_B$ being the Planck constant, speed of light, and Boltzmann constant, respectively. Population effects beyond the single temperature case have been explored in \cite{2022A&A...659A..70D} using a Gaussian $P(T)$. However, several arguments suggest that a lognormal power-law function is more physically motivated. On the low-temperature side, a lognormal distribution is preferred simply because temperature is strictly positively defined, whereas a Gaussian distribution allows unphysical negative temperatures. On the high-temperature end, a power-law tail is naturally expected, as it follows the same behavior as the interstellar radiation field (ISRF) that heats the dust \citep[][]{2001ApJ...549..215D,2007ApJ...657..810D}. Independent of our work, recent theoretical and observational studies have provided further support for a lognormal power-law distribution in describing dust temperature, dust emission, or the ISRF \citep{2024ApJ...975..173L,2024AJ....167...39P}. For our SED fitting, we adopt the modified lognormal power-law (MLP) function from \cite{2015MNRAS.449.2413B} for temperature $T$ in K:
\begin{eqnarray}
P(T)\, =\, && P(T,\, \mu_T,\, s_T,\, \alpha_T) \nonumber\\ \propto\, &&\frac{\alpha_T}{2}\, {\rm exp}\,({\alpha_T \mu_T + \alpha_T^2 s_T^2/2})\,\, T^{-(1+\alpha_T)}\,  \nonumber\\ &&\times\,{\rm erfc}\left(\frac{1}{\sqrt{2}}\,\left(\alpha_T s_T - \frac{{\rm ln}\,T-\mu_T}{s_T} \right) \right)\,,
\label{eq:MLP}
\end{eqnarray}
where $\rm erfc$ is the complementary error function, and $\mu_T$, $s_T$, and $\alpha_T$ are free parameters describing the characteristic logarithmic temperature, spread, and high-end power index, respectively. The MLP functional form may appear complicated at first glance, but it provides an elegant way to avoid manually stitching a lognormal and a power-law at an arbitrary transition point. 

We now describe the redshift dependence for each element in Equation~\ref{eq:SED_1}, allowing enough flexibility to include galaxy evolution effects. For the overall normalization tied to the dust mass density $\rho_{\rm d}$, we adopt a general functional form (but not the specific coefficients) from \cite{2014ARA&A..52..415M}, characterized by a redshift peak with free location, width, and independent slopes at low and high redshift:
\begin{eqnarray}
\rho_{\rm{d}}(z)\, =\, a\,\frac{(1+z)^b}{1+[(1+z)/c]^d}\ \ \rm{M_{\odot}\ Mpc^{-3}}\,, 
\label{eq:z_evo_norm}
\end{eqnarray}
with four free parameters: $a$, $b$, $c$, and $d$. While literature results suggest the presence of a peak at cosmic noon, this function also allows for monotonic evolution if supported by the data. For the Gaussian opacity spectral index distribution $P(\beta)$, we assume a free but non-evolving spread $\sigma_{\beta}$, and an evolving mean with two free-parameters, $\beta_0$, $C_{\beta}$, following:
\begin{eqnarray}
\langle \beta \rangle (z)\, =\, \beta_0 + C_{\beta}\, \textrm{log}(1+z)\,.
\label{eq:z_evo_beta}
\end{eqnarray}
For the MLP dust temperature distribution $P(T)$, we allow free but non-evolving $s_T$ and $\alpha_T$. The characteristic temperature, however, is allowed to evolve with redshift:  
\begin{eqnarray}
\mu_T(z)\, =\, \mu_0 + C_{\mu}\, \textrm{log}(1+z)\,.
\label{eq:z_evo_T}
\end{eqnarray}

\subsubsection{Very Cold Dust}\label{sec:cold_CGM}

Dust is found not only in the ISM but also in the CGM. The warm-cold component in Section~\ref{sec:model_warm_cold} could accommodate CGM dust if its temperature remains a continuation of the ISM dust distribution, thus the global temperature distribution remains unimodal. However, if dust survives for an extended period of time in cold clouds within the hot CGM or is transported further into the diffuse IGM, its thermal properties may be significantly altered. At galactic radii beyond several tens of kpc, radiative heating is no longer dominated by the local interstellar radiation field but instead governed by the metagalactic background \citep{2014ApJ...792....8W}. 

Unlike the commonly assumed temperature floor set by the CMB \citep{2013ApJ...766...13D}, a more precise picture is given by \cite{2016ApJ...825..130I}, who computed a characteristic temperature, here denoted as $T_{\rm floor}$, for dust in radiative equilibrium with the full EBL, accounting for both the CMB and a significant contribution from the UV background. By fitting their numerical results with a power law redshift evolution, we find this floor temperature well described by:
\begin{eqnarray}
T_{\rm floor}(z)\, = \, 4.292 \,(1+z)^{0.935}\, \textrm{K}\,,
\label{eq:T_cold}
\end{eqnarray}
shown by the dashed line in the right panel in Figure~\ref{fig:T_dust}. This temperature is lower than the typical temperature of ISM dust, particularly at $z=0$, and remains about 50\% higher than $T_{\rm CMB}$.  

So far, there is no direct observational evidence confirming that CGM dust predominantly exists in this very cold state. Consequently, its cosmological mass contribution remains uncertain. However, the lack of detections in the literature is consistent with the expected low surface brightness, given the already extended nature of the CGM and the steep $L \propto T^{4+\beta}$ scaling of dust emission, making observations extremely challenging.  

The wide wavelength range used in our analysis provides us with an opportunity to test the scenario of a hidden, cold dust reservoir in the universe. The additional temperature information complements analyses based on spatial correlations of dust reddening \citep[][]{2010MNRAS.405.1025M,2015ApJ...813....7P,2025arXiv250304098M} and helps break the degeneracy between three scenarios: (1) truly diffuse, very cold CGM/IGM dust and (2) warmer ISM dust in satellite galaxies with local heating, and (3) contribution from dust in more massive galaxies through two-halo clustering, all of which could produce similar spatial two-point functions.

To extract constraints on possible cold CGM dust using our densely sampled CIB spectrum, we consider an extended dust SED model including an additional ``very cold'' component,\footnote{Following the terminology in the literature for dust colder than 10~K, while labeled with the subscript ``cold'' in related parameters for brevity.} $\epsilon_{\nu,\, \rm{cold}}$, so that the total emissivity is:
\begin{eqnarray}
\epsilon_{\nu,\, \rm{total}} = \epsilon_{\nu} + \epsilon_{\nu,\, \rm{cold}}\,,
\label{eq:eps_nu_plus_cold}
\end{eqnarray}
where $\epsilon_{\nu}$ is the emissivity for the bulk warm-cold dust from Equation~\ref{eq:SED_1}. Assuming the optically thin case in Equation~\ref{eq:optical_thin_thick}, the emissivity of the very cold dust is given by:
\begin{eqnarray}
\epsilon_{\nu,\, \rm{cold}}\, &=&\, 4\pi\, \rho_{\rm{d,cold}}\, \kappa_0 \,\biggl \langle \left(\frac{\nu}{\nu_0}\right)^{\beta} \biggl \rangle \, B_{\nu}(T_{\rm floor})\,,
\label{eq:SED_CGM}
\end{eqnarray}
with a single temperature fixed to the evolving $T_{\rm floor}$ using Equation~\ref{eq:T_cold}. This effectively assumes that very cold CGM dust follows a delta function $P(T)$, and thus the combined $P(T)$ for all cosmic dust becomes bimodal. This provides a specific and critical test for possible very cold dust accumulating at the EBL heating floor, but we note that CGM and IGM dust does not need to be all in this state, and the warmer subset would have already been included in the main component allowing continuous and broad $P(T)$.

In Equation~\ref{eq:SED_CGM}, we adopt the same opacity anchor $\kappa_0 = 0.77~\rm cm^2\,g^{-1}$ and $\beta$ distribution $P(\beta)$ (for frequency dependence in the angle brackets through Equation~\ref{eq:tau_effective} and \ref{eq:P_beta}) as in the fiducial warm-cold dust model. To parameterize the normalization, and thus the comoving density of the very cold CGM dust $\rho_{\rm{d, cold}}$ (and $\Omega_{\rm{dust, cold}}$), we define a cold fraction $f_{\rm{cold}}$ such that:
\begin{eqnarray}
\rho_{\rm{d, cold}} = f_{\rm{cold}} \times \rho_{\rm{d}}\,,
\label{eq:rho_cold}
\end{eqnarray}
where $\rho_{\rm{d}}$ is the comoving density of the main component (Equation~\ref{eq:z_evo_norm}). We allow $f_{\rm{cold}}$ to evolve with redshift, following:
\begin{eqnarray}
f_{\rm{cold}}\, =\, f_{\rm{cold, 0}} \,(1+z)^{\gamma_{\rm {cold}}}\,,
\label{eq:z_evo_f_cold}
\end{eqnarray}
which completes the extended model with two extra parameters $f_{\rm{cold, 0}}$ and $\gamma_{\rm {cold}}$. Given the evolution of $T_{\rm floor}$, the contribution from very cold CGM dust is expected to peak at $\sim250$~GHz (920~GHz) in rest-frame emissivity at $z=0$ ($z=3$).

\subsubsection{Bayesian Inference and MCMC}\label{sec:MCMC}

We perform a Bayesian inference using the Markov Chain Monte Carlo (MCMC) method for SED fitting. The Bayes' theorem is given by:
\begin{equation}
  \mathrm{P}(\boldsymbol{\theta}\,|\textbf{\emph{D}}) \propto \mathrm{P}(\textbf{\emph{D}}\,|\,\boldsymbol{\theta})\, \mathrm{P}(\boldsymbol{\theta})\;, 
\end{equation}
where $\boldsymbol{D}$ and $\boldsymbol{\theta}$ are the data and model parameter vectors, respectively, $\mathrm{P}(\boldsymbol{\theta}\,|\textbf{\emph{D}})$ is the posterior, $\mathrm{P}(\textbf{\emph{D}}\,|\,\boldsymbol{\theta}) = \mathrm{P}(\textbf{\emph{D}}\,|\,\boldsymbol{\theta},\ \textbf{\emph{M}})$ is the likelihood function $L$ given model $\textbf{\emph{M}} = \textbf{\emph{M}}(\boldsymbol{\theta})$, and $\mathrm{P}(\boldsymbol{\theta})$ is the prior. Below, we specify each component.

\begin{itemize}[leftmargin=*]
\item Data vector $\textbf{\emph{D}}$: The main component is the tomographic $\epsilon_{\nu}b$ across frequency and redshift (Figure~\ref{fig:rest_spectrum}). Out of the original 176 data points, we use only 157 with rest frequencies below 8~THz (37.5~$\mu$m). This cut is to exclude polycyclic aromatic hydrocarbon (PAH) features in the mid-infrared EBL, which are beyond the scope of this paper. For tentative detections of the cosmic PAH background, we refer the readers to \cite{2019ApJ...870..120C} and \cite{2024ApJ...960...96C}. The 37.5~$\mu$m cut also excludes AGN contributions in the mid-infrared \citep[][and references therein]{2018MNRAS.478.4238D, 2022Univ....8..304L}, thus simplifying our interpretation of star-formation-driven energy input in the CIB. The classic single-temperature MBB cannot fit dusty galaxy or CIB SEDs at wavelengths below $\sim 80$~$\mu$m, while our generalized MBB with a temperature distribution naturally reproduces the mid-infrared power-law behavior similar to \cite{2012MNRAS.425.3094C}, but offering a more physical interpretation than phenomenological parameterization.

\begin{table*}[t!]
\begin{center}
%\centering
\caption{\label{table:parameters} \normalsize Priors and Posteriors of Model Parameters}
\renewcommand{\arraystretch}{1.1}
\hspace{-1.cm}
\begin{tabular}{llrll}
\hline 
Parameter & Meaning & \multicolumn{2}{l}{$\ \ \ \ \ $Range and Prior} & Posterior \\ 
\hline
log~$a$\,$^{\mathtt{a}}$ & Dust density normalization & [4, 6] & flat & $5.16^{+0.11}_{-0.11}$ \\ 
$b$ & Dust density evolution & [0, 4] & flat & $1.54^{+0.32}_{-0.31}$ \\
 $c$ & Dust density evolution & [0, 5] & flat & $2.83^{+0.14}_{-0.14}$ \\ 
 $d$ & Dust density evolution & [4, 9] & flat & $6.55^{+0.40}_{-0.35}$\\ 
 $\mu_0$ & Characteristic logarithmic dust temperature at $z=0$ & [2, 3] & flat & $2.55^{+0.09}_{-0.08}$\\ 
 $C_{\mu}$ & Characteristic logarithmic dust temperature evolution & [$-1$, 2] & flat & $0.60^{+0.09}_{-0.09}$\\ 
 $s_T$ & Logarithmic dust temperature intrinsic spread & [0, 0.2] & flat & $0.04^{+0.05}_{-0.03}$\\ 
 $\alpha_T$ & Dust temperature distribution high-end power index  & [4, 6] & flat & $5.00^{+0.21}_{-0.16}$\\ 
 log~$\Sigma_{\rm d}$\,$^{\mathtt{a}}$ & Effective dust surface density for optically thick transition & [7, 8.5] & $\mathcal{N}(7.7,\,0.1)$ & $7.78^{+0.09}_{-0.09}$\\ 
 $\beta_0$ & Mean dust opacity spectral index at $z=0$ & [1.3, 3] & flat & $1.69^{+0.12}_{-0.12}$\\ 
 $C_{\beta}$ & Mean dust opacity spectral index evolution & [$-1.5$, 1.5] & flat & $0.37^{+0.19}_{-0.18}$\\ 
 $\sigma_{\beta}$ & Intrinsic spread for dust opacity spectral index & [0.1, 1.5] & flat & $0.71^{+0.11}_{-0.12}$\\ 
 \hline
 $b_0$ & Effective CIB bias at $z=0$ & [0, 1.5]\,$^{\mathtt{b}}$ & $\mathcal{N}(1,\,0.1)$ &  $0.94^{+0.10}_{-0.10}$\\ 
 $b_1$ & Effective CIB bias evolution & [0.5, 1.5]\,$^{\mathtt{b}}$ & flat & $0.66^{+0.08}_{-0.09}$\\ 
 $b_2$ & Effective CIB bias evolution & [0, 1]\,$^{\mathtt{b}}$ & flat & $0.04^{+0.05}_{-0.03}$\\ 
  \hline
  \hline
 $f_{\rm{cold,\, 0}}\,^{\mathtt{c}}$ & Very cold dust density fraction at $z=0$ & [0, 2] & flat & $< 1.56$\\ 
 $\gamma_{\rm {cold}}\,^{\mathtt{c}}$ & Very cold dust density fraction evolution & [$-4$, 1] & flat & $< -1.10$\\ 
\hline
\vspace{-0.3cm}\\
\multicolumn{4}{p{13cm}}{$^{\mathtt{a}}$ We use logarithmic forms for $a$ and $\Sigma_{\rm d}$ in the fitting.}\\
\multicolumn{4}{p{13cm}}{$^{\mathtt{b}}$ Additional bounds, $b_{z=2} > 2.4$ and $b_{z=3} < 4.81$, are set for the combined bias factor in Equation~\ref{eq:b_of_z}.}\\
\multicolumn{4}{p{13cm}}{$^{\mathtt{c}}$ Very cold dust is added in a separate MCMC run; as it is undetected, we report $1\sigma$ upper limits.}
\vspace{-0.5cm}\\
\end{tabular}
\end{center}
\end{table*}

\begin{figure*}[t!]
    \begin{center}
         \includegraphics[width=0.85\textwidth]{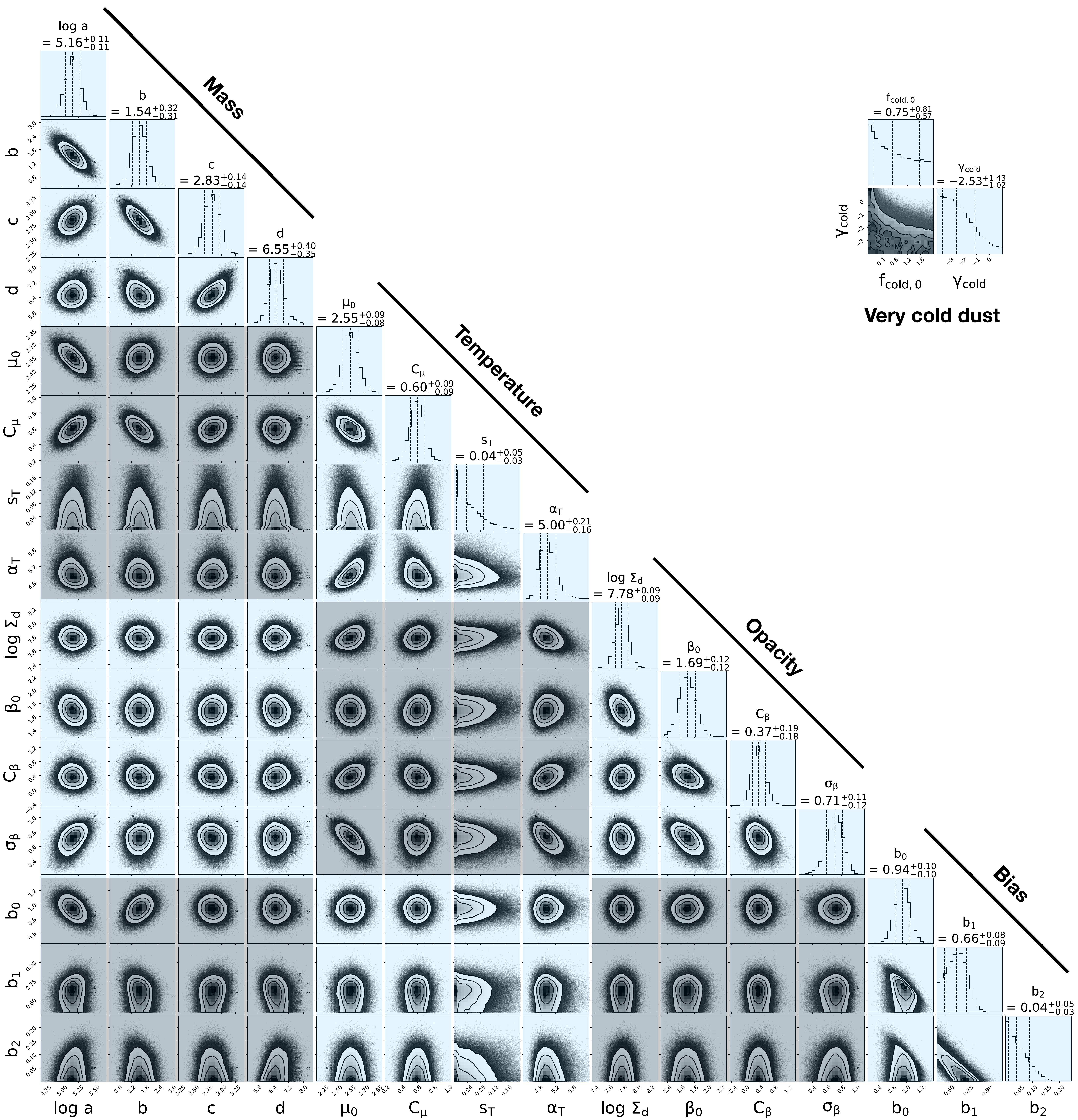}
    \end{center}
    \caption{Marginalized 1D posteriors and pairwise 2D covariances for the 15 parameters in our fiducial warm-cold ensemble cosmic dust SED model fit to the tomographic CIB measurements. The physical role of the parameters can be organized into dust mass, temperature, opacity, and bias, and only three parameters are prior-driven (log~$\Sigma_{\rm d}$ and two DOF for the bias). We include very cold CGM/IGM dust in a separate MCMC run, with posteriors for the two additional parameters shown in the upper right. Since this extra cold component is not significantly detected, the posteriors for the remaining 15 parameters in this run remain consistent with the fiducial results shown in the main block.}
    \label{fig:posterior}
\end{figure*}

To break the bias-emissivity degeneracy, we include the CIB monopoles, i.e., the redshift-integrated total intensities, in the data vector (and their errors in the covariance matrix used for the Likelihood). We use the monopole measurements from \cite{2019ApJ...877...40O} for each {\it Planck} HFI channel, which combines {\it Planck} data with measurements from the Far Infrared Absolute Spectrophotometer (FIRAS) on the Cosmic Background Explorer \citep{1998ApJ...508..123F}. Although FIRAS is absolutely calibrated, at frequencies above $\sim$600~GHz, the estimated monopole becomes self-inconsistent when different methods are used to mitigate the Milky Way thermal dust foreground. We therefore include the monopole only up to 545~GHz (5 bands). In total, our data vector consists of 162 elements:
    \begin{eqnarray}
    \textbf{\emph{D}}\, =\, (&&\epsilon_{\nu}b_1,\ \epsilon_{\nu}b_2,\ \epsilon_{\nu}b_3,\ ...,\ \epsilon_{\nu}b_{157},\ \nonumber \\ && I_{\nu}^{100}, \ 
    I_{\nu}^{143}, \ I_{\nu}^{217}, \ I_{\nu}^{353}, \ I_{\nu}^{545}\,) \,.
     \label{eq:data_vector}
    \end{eqnarray}

\item Model $\textbf{\emph{M}}$: Given our data space, the model must describe two quantities: the tomographic $\epsilon_{\nu}b$ at any frequency and redshift, and the monopole $I_{\nu}$ at any observed band. Using the realistic ensemble cosmic dust SED in Section~\ref{sec:model}, the bias-weighted emissivity is simply the product:
    \begin{eqnarray}
    \epsilon_{\nu}b(\nu, z)\, &=& \, \epsilon_{\nu}(\nu, z) \, b(z)\,,
    \label{eq:model_warm_cold_only}
    \end{eqnarray}
where the emissivity term is from Equation~\ref{eq:SED_1} for the fiducial warm-cold dust scenario or Equation~\ref{eq:eps_nu_plus_cold} for the extended scenario with very cold dust, and the bias term is from Equation~\ref{eq:b_of_z}.

For the monopole at an observed frequency $\nu_{\rm obs}$, we first convert the model $\epsilon_{\nu}$ at rest frequency $\nu = (1+z)\,\nu_{\rm obs}$ into the redshift differential intensity $dI_{\nu}/dz$ following Equation~\ref{eq:dIdz_to_emissivity}, without multiplying by $b$. The monopole is then given simply by the integral:
    \begin{eqnarray}
    I_{\nu}(\nu_{\rm obs})\, = \, \int \frac{dI_{\nu}}{dz}(\nu, z)\, dz\,,
    \label{eq:monopole}
    \end{eqnarray}
with the integration bounds set from $z=0$ to 10. 

During the fitting stage, we compute the model $\epsilon_{\nu}b$ at the 157 ($z$, $\nu$) sampling points and $I_{\nu}$ at the 5 $\nu_{\rm obs}$ values to compare with our data vector $\textbf{\emph{D}}$. 

In total, 15 free parameters are required in our fiducial model $\textbf{\emph{M}}$ to describe the evolving CIB spectrum, as summarized in Table~\ref{table:parameters}. These include 12 parameters for the ensemble warm-cold dust thermal continuum ($a$, $b$, $c$, $d$, $\mu_0$, $C_{\mu}$, $s_T$, $\alpha_T$, $\Sigma_{\rm d}$, $\beta_0$, $C_{\beta}$, $\sigma_{\beta}$) and 3 for the effective bias ($b_0$, $b_1$, $b_2$). In the extended model allowing very cold dust as a second mode in the temperature distribution, 2 additional parameters ($f_{\rm{cold, 0}}$, $\gamma_{\rm {cold}}$) are needed.

\item Likelihood $L = \mathrm{P}(\textbf{\emph{D}}\,|\,\boldsymbol{\theta},\ \textbf{\emph{M}})$: We assume a Gaussian likelihood with covariances, where the log-likelihood is:
    \begin{eqnarray}
    \textrm{log}\,L\, \propto \, -\frac{1}{2}\,\Delta \textbf{\emph{D}}^T\,\bm{C}^{-1}\, \Delta \textbf{\emph{D}}\,,
    \label{eq:log_likelihood}
    \end{eqnarray}
where $\Delta\textbf {\emph{D}}$ is the difference between the data and model vectors, and $\textbf{\emph{C}}$ is the covariance matrix. We use the empirical covariances for $\epsilon_{\nu}b$ from Figure~\ref{fig:covariance}, trimmed below the maximum rest frequency (8~THz). Since the monopoles $I_{\nu}$ are appended after $\epsilon_{\nu}b$ in $\textbf{\emph{D}}$, we extend the covariance matrix by incorporating the per-band monopole uncertainties from \cite{2019ApJ...877...40O}, assuming no cross-band covariance for the additional off-diagonal elements.

\item Prior $P(\boldsymbol{\theta})$: For all but one spectral parameter, we use flat priors with broad, uninformative bounds, keeping the constraints maximally empirical. The only exception is a Gaussian prior for the galaxy-scale dust surface density: $\log \Sigma_{\rm d}/(\rm M_{\odot}\ kpc^{-2}) \sim \mathcal{N}(7.7,\,0.1)$. This is chosen such that the transition wavelength $\lambda_{\rm thick}$ between the optically thin and thick regimes is around $60$--$80\,\mu$m for typical $\beta \sim 2$ (see Equation~\ref{eq:lambda_thick}), consistent with the results in \cite{2017ApJ...839...58S}. 

For the bias factor $b(z)$ with three DOF, $b_0$, $b_1$, and $b_2$ (Equation~\ref{eq:b_of_z}), our data vector, combining clustering tomography with bias-free monopoles, constrains only one effective amplitude. The redshift evolution must therefore be prior-driven. We impose a Gaussian prior on $b_0 \sim \mathcal{N}(1,\,0.1)$ at $z=0$, consistent with the bias of dark matter halos in the mass range M$_h \sim 10^{12}$--$10^{13}$~M$_{\odot}$ \citep{2010ApJ...724..878T}. At cosmic noon, dusty galaxies are generally more clustered than unobscured star-forming populations \citep[e.g.,][]{2020ApJ...904....2G}. We thus enforce $b(z=2) > 2.4$ as a lower bound, corresponding to the bias of Lyman break galaxies \citep[][]{2005ApJ...619..697A}. Finally, a theoretical upper limit can be imposed: At all redshifts, the CIB bias should not exceed $b_y$, the effective bias for the SZ effect Compton-$y$ signal from massive clusters and protoclusters. In a halo model framework, Compton-$y$ scales super-linearly with halo mass as $y \propto M_{\rm h}^{5/3}$. In contrast, the far-infrared light likely scales with halo mass sub-linearly at the massive end \citep[e.g.,][]{2012A&A...537L...5B, 2018MNRAS.475.3974W}, making the CIB not as biased. This constraint is most restrictive at high redshift, so we impose $b < b_y = 4.81$ at $z=3$, where $b_y$ is taken from the halo model in \citetalias{2020ApJ...902...56C}. 

We summarize the priors described above in Table~\ref{table:parameters}.

\end{itemize}

To obtain the posterior distributions, we perform MCMC sampling using an affine-invariant ensemble sampler implemented in the {\it emcee} code \citep{2013PASP..125..306F}. We carry out two MCMC runs, first with the fiducial warm-cold dust model, and second with the extended scenario allowing very cold dust. 

\begin{figure}[t!]
    \begin{center}
         \includegraphics[width=0.465\textwidth]{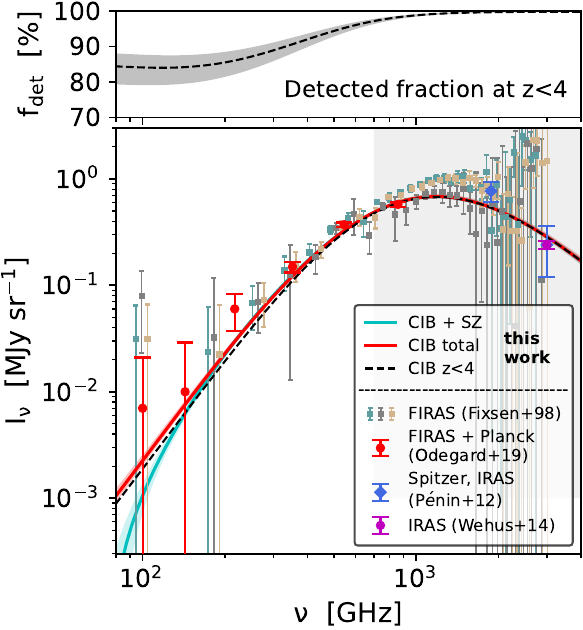}
    \end{center}
    \caption{Observer-frame CIB monopole spectrum. The red line and band show the posterior median and $68\%$ range from our CIB tomography. The black dashed line represents the $z<4$ contribution resolved directly via cross-correlations, with its ratio to the total shown in the top panel. The cyan line and band show the CIB plus the SZ effect of the CMB due to hot gas in the LSS. Data points represent constraints from the literature, among which the \cite{2019ApJ...877...40O} points up to $545$~GHz are used in our MCMC fitting (so our monopole estimate is not an independent measurement). Our high-frequency posterior suggests that the FIRAS monopole cleaned with HI (gray squares) is the most robust among the three foreground-cleaning methods in \cite{1998ApJ...508..123F}.}
    \label{fig:monopole}
\end{figure}

\subsubsection{Goodness of Fit and Posteriors}\label{sec:sed_fit_result}

We show in Figure~\ref{fig:posterior} the marginalized 1D posteriors and 2D covariances for the 15 parameters in the fiducial fit with the ensemble warm-cold cosmic dust model. Most parameters exhibit well-localized peaks in their posterior. The best-fit model is shown in Figures~\ref{fig:dIdzb} and \ref{fig:rest_spectrum} with a set of lines that closely follow the data points, yielding a reduced $\chi^2$ of 1.49. This calculation includes a correction for inverting a noisy covariance matrix following \cite{2007A&A...464..399H}.

For comparison, fitting a simpler model with a single-temperature MBB with evolving $T$ and $\beta$ under general opacity assumption results in poor goodness of fit, with a reduced $\chi^2$ of 2.76. We further test a case extending the high-frequency bound of the data vector from 8~THz to 10~THz, slightly deeper into the mid-infrared. In this case, the reduced $\chi^2$ for the single-temperature MBB model worsens significantly to 3.77, whereas our fiducial ensemble cosmic dust model continues to provide a good fit with no change in reduced $\chi^2$. This demonstrates that the complexity of the CIB spectrum extends far beyond a single MBB component, as hotter dust exists in locally heated regions across the diverse galaxy populations that make up the CIB.

In the extended MCMC run with very cold dust added, the full 17-parameter posterior is obtained. The 2D posterior distribution for the two additional cold dust parameters is shown in the upper-right corner of Figure~\ref{fig:posterior}. Unfortunately, despite the attempt and our dense spectral sampling, very cold dust is not detected (best-fit $f_{\rm cold}$ in Equation~\ref{eq:z_evo_f_cold} not significantly greater than zero). CGM/IGM dust, thus, may not be primarily in this thermal state and have already been captured in our warm-cold component. This null detection also implies that the posteriors for the remaining 15 parameters remain largely unchanged from the fiducial warm-cold-only fit, as already presented in Table~\ref{table:parameters} and the main block in Figure~\ref{fig:posterior}. Given the insignificance of very cold dust, from what follows, we discuss the result mainly for the fiducial warm-cold fit.

In Figure~\ref{fig:monopole}, we show our posterior CIB monopole as the red line (also listed as $I_{\nu}^{\ \mathtt{e}}$ in Table~\ref{table:per_band_stats}), with the red-shaded region representing the $68\%$ confidence range. This is not an independent measurement but primarily constrained by the direct monopole measurements (red data points) from \cite{2019ApJ...877...40O} at and below $545$~GHz and, indirectly, by our multi-band redshift tomography above $545$~GHz.  

For comparison, in Figure~\ref{fig:monopole}, we overlay the FIRAS CIB monopole from \cite{1998ApJ...508..123F}, where foregrounds were mitigated using three different methods, with results diverging at frequencies above 1~THz. Two of these methods use far-infrared foreground templates, while the third relies on neutral hydrogen HI. Our best fit at high frequencies aligns more closely with the one cleaned by HI (gray squares), with the other two FIRAS results under-subtracting the foregrounds (teal and tan squares). This highlights the difficulty of far-infrared component separation unless external data with redshift or velocity information is incorporated. Figure~\ref{fig:monopole} also compares our best fit  with measurements from \cite{2012A&A...543A.123P} and \cite{2014arXiv1411.7616W}, showing a more complicated picture at high frequencies \citep[see also][]{1998ApJ...508...25H, 2000A&A...355...17L, 2012A&A...542A..58B}. 

In Figure~\ref{fig:monopole}, the black dashed line represents the portion of the CIB monopole we directly resolve or detect up to $z=4$ through tomographic cross-correlations (Equation~\ref{eq:monopole}). As shown in the top panel, the resolved fraction at $z<4$ is nearly $100\%$ at high frequencies and remains above $80\%$ at low frequencies, primarily due to the negative $K$-correction shifting the peak $dI_{\nu}/dz$ to higher redshifts.

\begin{figure}[t!]
    \begin{center}
    \includegraphics[width=0.47\textwidth]{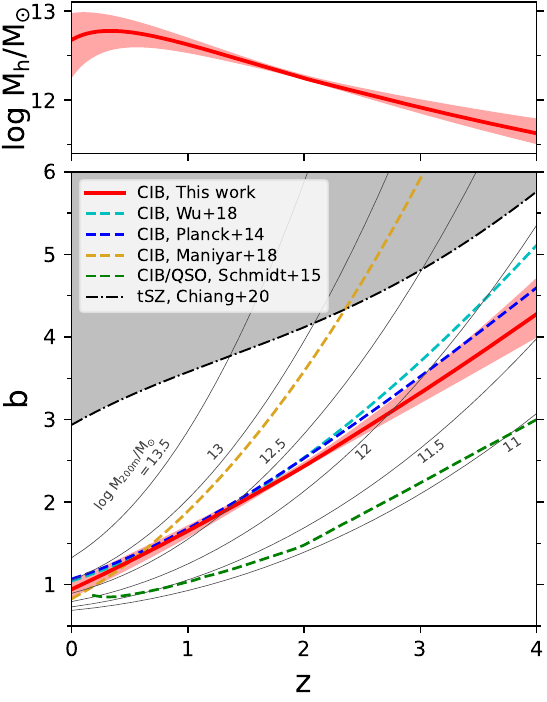}
    \end{center}
    \caption{Effective CIB bias evolution. Red line and shaded region show the posterior median and 68$\%$ range, constrained empirically by combining clustering-based tomography with FIRAS-{\it Planck}-based monopole intensities in Figure~\ref{fig:monopole}. Colored dashed lines represent literature constraints. The effective thermal SZ bias (dashed-dotted line) from the halo model in \citetalias{2020ApJ...902...56C} provides a theoretical upper limit to the CIB bias. Thin gray lines indicate dark matter halo bias from \cite{2010ApJ...724..878T}, with the effective light-weighted mean dark matter halo mass for the CIB shown in the upper panel.}
    \label{fig:bias}
\end{figure}

As our primary interest lies in astrophysical information, the effective CIB bias factor $b$ can be considered a nuisance parameter. However, it still carries valuable insights into the connection between dark matter halos and the galaxies contributing to the CIB. Our best-fit CIB bias is 
\begin{eqnarray}
b(z) = 0.94 + 0.66\,z + 0.04\,z^2 \,,
\label{eq:b_of_z_best_fit}
\end{eqnarray}
which we show in Figure~\ref{fig:bias} with the red line and shaded region, alongside other literature constraints.

Our posterior $b(z)$, constrained by redshift tomography combined with the FIRAS-{\it Planck} monopole, is consistent with \cite{2014A&A...571A..30P} and \cite{2018MNRAS.475.3974W}, who fit the {\it Planck} CIB auto power spectra using halo models of varying complexity. In contrast, the CIB bias from \cite{2018A&A...614A..39M}, derived using CIB auto power spectra and 
CIB-CMB lensing, deviates and exceeds a theoretical bound set by the SZ effect bias $b_y$ (dashed-dotted line). Explaining such a high bias in the halo model framework would require an effective far-infrared–halo mass scaling steeper than $M_{\rm h}^{5/3}$, difficult to sustain given the strong CIB amplitudes and the rarity of massive halos. We also compare our results with the CIB bias constraints from \cite{2015MNRAS.446.2696S}, obtained using tomographic QSO-CIB cross-correlations and CIB auto-correlations. This measurement is lower than the rest, possibly due to systematics they encounter, as discussed in Section~\ref{sec:result_CIB_spectrum}. 

We caution that our posterior $b(z)$ may be underestimated if the cross-correlation coefficient, defined as $r = \overline{w}_{\rm{IR}} / \sqrt{\overline{w}_{\rm{II}} \, \overline{w}_{\rm{RR}}}$ with $\rm I$ being the (deprojected) CIB and $\rm R$ the SDSS references, is below unity. In that case, Figure~\ref{fig:bias} should be interpreted as showing $b \times r(z)$. Uncertainty in $r$ likely dominates the systematics in our $b(z)$ estimates, and the errors shown reflect only statistical uncertainty. However, this does not affect the dust mass and cosmic star formation estimates in later sections to first order, as the intensity normalization is anchored to the {\it Planck} + FIRAS CIB monopole.

Assuming $r=1$, Figure~\ref{fig:bias} also compares our CIB bias with the dark matter halo bias from \cite{2010ApJ...724..878T} as shown by the thin gray lines for our assumed cosmology. This provides an estimate for $M_{\rm h, CIB}$, the light-weighted mean halo mass associated with the CIB, displayed in the upper panel. We find that $M_{\rm h, CIB}$ increases from $\sim 10^{12.0}~\rm M_{\odot}$ at $z=3$ to $10^{12.7}~\rm M_{\odot}$ at $z=0$. This evolution is milder than that of the nonlinear mass scale $M_{*}$ for halo formation \citep{1974ApJ...187..425P}, which grows rapidly from $10^{8}~\rm M_{\odot}$ at $z=3$ to $10^{12.5}~\rm M_{\odot}$ at $z=0$. Thus, we observe a form of cosmic ``downsizing'' for the CIB, where $M_{\rm h, CIB}/M_*$ decreases over time, though the trend is not as strong as found in \cite{2018MNRAS.480..149H}.

The $b(z)$ in Figure~\ref{fig:bias}, combined with $(dI_{\nu}/dz)\,b$ from Figure~\ref{fig:dIdzb} allows us to derive an alternative set of CIB monopole estimates in Table~\ref{table:per_band_stats}. This is done by simply dividing out $b(z)$ for all $(dI_{\nu}/dz)\,b$ data points and computing a direct summation, band by band, for the integral $I_{\nu} = \int (dI_{\nu}/dz)\, dz$. These monopoles up to $z=4$ and those with small extrapolations to $z=\infty$ (using $f_{\rm det}$ in Figure~\ref{fig:monopole}) are tabulated as $I_{\nu}^{\ \mathtt{c}}$ and $I_{\nu}^{\ \mathtt{d}}$ in Table~\ref{table:per_band_stats}. We refer to these $I_{\nu}^{\ \mathtt{d}}$ as the direct summation monopoles and those in $I_{\nu}^{\ \mathtt{e}}$ as the best-fit model monopoles, and the two are consistent within errors.

\begin{figure*}[t!]
    \begin{center}
    \includegraphics[width=\textwidth]{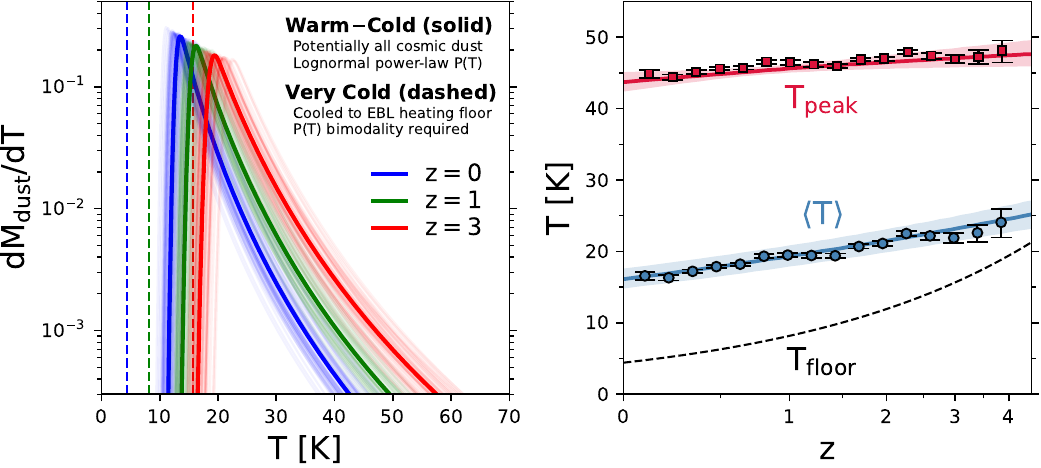}
    \end{center}
    \caption{Dust temperature constraints from our CIB tomography. Left: Full $P(T) \sim dM_{\rm{dust}}/dT$ distributions, with thick lines for posterior medians and thin lines for selected posterior samples at $z=0$, 1, and 3. Right: Evolution of the mass-weighted mean $\langle T \rangle$ (blue) and emissivity peak temperature $T_{\rm peak}$ (red), indicating stronger interstellar dust heating at higher redshifts. Dashed lines in both panels represent a theoretical temperature floor set by the metagalactic radiation background. At a given epoch, the main lognormal power-law component could, in principle, capture all dust regardless of the local environment, e.g., ISM, CGM, or IGM, as long as $P(T)$ is single-modal. If the secondary cold peak at $T_{\rm floor}$ is significant, as explored in Section~\ref{sec:cold_CGM}, more dust mass could be allowed given the observed CIB, but only at $z\sim 0$, where $T_{\rm floor}$ is low and the temperature gap is large.}
    \label{fig:T_dust}
\end{figure*}

\subsection{Cosmic Dust}\label{sec:T_dust_Omega_dust}

We now examine the astrophysical information on cosmic dust, focusing on its temperature and mass budget.

\subsubsection{Temperature}\label{sec:T_dust}

For the first time, we constrain a realistic temperature distribution $P(T)$ for the ensemble cosmic dust and its evolution over a wide redshift range. From Equations~\ref{eq:SED_1}, \ref{eq:optical_thin_thick}, and \ref{eq:B_effective}, it can be appreciated that $P(T)$ is mass-weighted, i.e., $P(T) \propto (dM_{\rm dust}/dT)(T)$ in the optically thin regime. This remains a good approximation in our general opacity case, as deviations only occur at the highest frequencies or for the highest temperatures of interest.  

In Figure~\ref{fig:T_dust} (left panel), we show the normalized $P(T) = dM_{\rm dust}/dT$ posterior median in thick lines at $z = 0$, 1, and 3 in blue, green, and red, respectively. Some posterior samples are shown in thin lines to illustrate the range of uncertainty. At a given epoch, dust temperatures follow an MLP distribution (Equation~\ref{eq:MLP}) with two key features: (1) a lognormal core peaking around 10--25 K, primarily constrained by the shape of the CIB SED near its peak at 1--5 THz in Figure~\ref{fig:rest_spectrum}, and (2) a power-law tail in $P(T)$ that drives the power-law behavior of the CIB SED into the mid-infrared, which significantly deviates from any single-temperature spectrum declining exponentially on the Wien side. Part of this high-temperature tail may originate from small dust grains that are stochastically heated and not necessarily in thermal equilibrium.

Given the full distribution, there could be any number of ways to quantify a characteristic temperature. Here, we focus on two. First, we consider the mass-weighted mean:
\begin{eqnarray}
\langle T \rangle = \frac{1}{M_{\rm dust}}\int \frac{dM_{\rm dust}}{dT}\, T\, dT\,,
\label{eq:T_mean}
\end{eqnarray}
which is simply the first moment of $P(T)$ in Figure~\ref{fig:T_dust} (left). Since the MLP shape is physically motivated and its free parameters are fully constrained by the data, our $\langle T \rangle$ is robust against simplistic SED assumptions and can be directly compared with simulations or theoretical predictions.  

Alternatively, one can define a characteristic temperature using the peak temperature $T_{\rm peak}$, determined via Wien's displacement law:
\begin{eqnarray}
T_{\rm peak} = \frac{\nu_{\rm peak}/\rm GHz}{58.79}\, \rm K\,,
\label{eq:T_peak}
\end{eqnarray}
where $\nu_{\rm peak}$ is the frequency at the maximal emissivity $\epsilon_{\nu}$. $T_{\rm peak}$ characterizes the shape of the SED and thus, roughly speaking, serves as a light-weighted temperature. We do not use the per-wavelength version defined at peak $\epsilon_{\lambda}$, often used for individual dusty galaxies. This is because the CIB exhibits a nearly flat $\epsilon_{\lambda}$ at high frequencies; thus, the peak is ill-defined, likely due to population effects.
%\begin{eqnarray}
%T_{\rm peak} = \frac{2898}{\lambda_{\rm peak}/\mu \rm m}\, \rm K\,,
%\label{eq:T_peak}
%\end{eqnarray}

In the right panel of Figure~\ref{fig:T_dust}, we present the redshift evolution of both $\langle T \rangle$ and $T_{\rm peak}$ in blue and red, respectively, with the lines and shaded bands showing the posterior medians and 68\% confidence ranges. The possible shapes of $\langle T \rangle(z)$ and $T_{\rm peak}(z)$ evolution are ultimately driven by our parameterization for $\mu_T(z)$ in Equation~\ref{eq:z_evo_T}. To validate this, we extract another set of constraints that are non-parametric in redshift. We do so via rerunning our MCMC SED fitting with 16 free $\mu_T$ parameters in $P(T)$ (Equation~\ref{eq:MLP}), one per redshift bin while keeping all other SED parameters fixed at the posterior median from our fiducial run. The resulting data points, shown in corresponding colors (and listed in Table~\ref{table:omega_dust_SFD_history}), follow the smooth evolution closely, supporting the validity of our redshift parameterization for $\mu_T$.  

At any given cosmic time, the light-weighted dust temperature, $T_{\rm peak}$, is higher than the mass-weighted temperature, $\langle T \rangle$, as expected due to the lower emissivity of colder but more mass-bearing dust. Both temperatures increase with redshift, consistent with the picture that high-redshift galaxies often have higher star formation rates (SFR), specific SFRs, and/or SFR surface densities, likely driven by higher cosmological gas inflow, and result in stronger ISRFs in the ISM \citep[][]{2019MNRAS.489.1397L}. The posterior temperature evolutions are well described by power laws:
\begin{eqnarray}
\langle T \rangle(z) &=& 16.1 \times (1+z)^{0.26} \, \rm K\,,\\
T_{\rm peak}(z) &=& 44.0 \times (1+z)^{0.05} \, \rm K\,,
\label{eq:T_dust_z_approx}
\end{eqnarray}
where the slope of $T_{\rm peak}(z)$ is somewhat shallower than that of $\langle T \rangle(z)$. Although a careful comparison with the literature requires homogenizing temperature definitions, we find that the CIB-based temperature for cosmic dust evolves more slowly than individual bright dusty galaxies found in \cite{2018A&A...609A..30S}, \cite{2020ApJ...902..112B}, \cite{2020MNRAS.498.4192F}, and \cite{2022MNRAS.516L..30V}, and is more consistent with the milder evolution found in \cite{2014A&A...561A..86M} and \cite{2022ApJ...930..142D},  and less extreme conditions extrapolating to higher redshifts \cite[e.g.,][]{2024MNRAS.527.6867A}. We speculate that some discrepancy might be due to selection effects in galaxy surveys, while our CIB-based result should be interpreted as the cosmic mean. 

Beyond the main body of warm-cold dust with broad MLP temperature distributions, Figure~\ref{fig:T_dust} also marks the expectation for the second component of very cold dust with $T_{\rm floor}$ from Section~\ref{sec:cold_CGM} and Equation~\ref{eq:T_cold} in vertical and evolving dashed lines in the left and right panels. As already mentioned, our data suggest that this component is insignificant, and the full temperature distribution of cosmic dust likely shows no bimodality.

\begin{figure*}[ht]
\centering
P\includegraphics[width=0.765\textwidth]{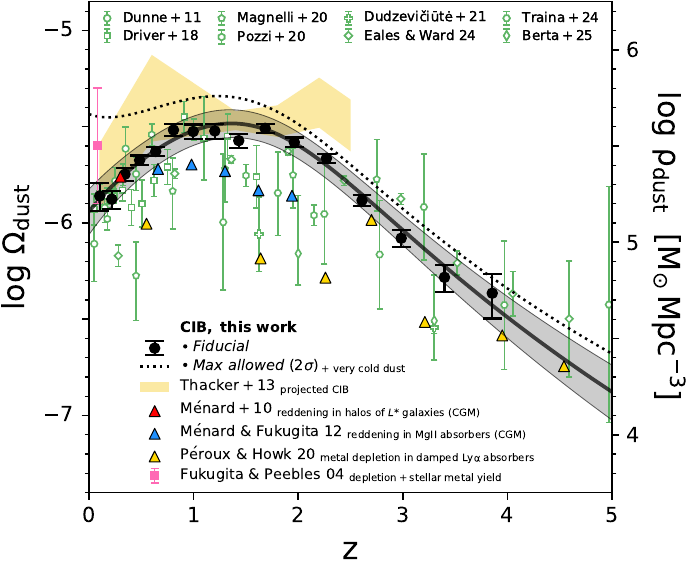}
\caption{Cosmic dust density $\rm \Omega_{dust}$ history. Black solid line and gray band show the posterior median and 68\% range from our CIB tomography, with black data points representing our per-redshift fits. Black dotted line marks an absolute upper limit (95\%) if temperature bimodality is allowed for additional very cold dust (Figures~\ref{fig:T_dust}). Various literature measurements and lower limits are shown as labeled, with green open symbols indicating those from far-infrared galaxy surveys. Most estimates, including far-infrared emission and optical reddening ones, are rescaled to assume the same reference opacity $\kappa_{850} = 0.77~\rm cm^2\,g^{-1}$. All errors shown, including ours, are statistical only. We find $\rm \Omega_{dust}$ increased rapidly in the first few Gyr, peaked at $z=1$--1.5, and declined at $z<1$. Note that our overall normalization is empirically constrained by the CIB monopole, and the uncertainty shown already includes that from the bias evolution (Figure~\ref{fig:bias}).}
\label{fig:Omega_dust}
\end{figure*}

Unlike single-component MBB fitting, our temperature constraint does not degenerate with that for the opacity or emissivity spectral index $\beta$ as shown in the posteriors in Figure~\ref{fig:posterior}. This is thanks to our full frequency coverage and the expanded parameterization for both $T$ and $\beta$ as realistic distributions. For completeness, we report the best-fit $\langle \beta \rangle$, the mean for the Gaussian $P(\beta)$ distribution in Equation~\ref{eq:P_beta}:
\begin{eqnarray}
\langle \beta \rangle (z)\, =\, 1.69 + 0.37\, \textrm{log}(1+z)\,,
\label{eq:beta_posterior}
\end{eqnarray}
which increases moderately with redshift.\footnote{We also find a large intrinsic scatter $\sigma_{\beta} = 0.7$, but degenerate with potential CO line contributions, which will be explored in a follow-up paper.} A lower $\beta$ could be associated with larger grain sizes \citep{2006ApJ...636.1114D}, so the mild evolution of $\langle \beta \rangle$ suggests slightly larger mean grain sizes for cosmic dust at late times, consistent with continued grain growth in the ISM over cosmic time.

\subsubsection{Mass Budget}\label{sec:Omega_dust}

One key piece of astrophysical information we aim to extract from the CIB deprojection is $\Omega_{\rm dust}(z)$, the comoving dust mass density expressed in units of the critical density, which enters through Equations~\ref{eq:SED_1} and \ref{eq:omega_dust}. $\Omega_{\rm dust}$ describes a key component of the cosmic inventory \citep{2004ApJ...616..643F} and serves as a valuable constraint for understanding and modeling the dust life cycle. Although many parameters are used to describe the full shape and evolution of the CIB spectrum, the estimate of dust mass is relatively independent of this modeling, as it scales directly with the low-frequency amplitude observed in the optically thin Rayleigh-Jeans regime. In Figure~\ref{fig:Omega_dust}, we present our constraints on the redshift evolution of $\Omega_{\rm dust}$, with the black line and gray shaded band representing the posterior median and 68\% range, respectively. We also extract per-redshift $\Omega_{\rm dust}$ posteriors, shown as black data points in Figure~\ref{fig:Omega_dust} and listed in Table~\ref{table:omega_dust_SFD_history}. To be non-parametric in redshift evolution, this is done by refitting the tomographic SED using 16 free $\rho_{\rm d}$, one for each redshift bin while fixing all other spectral shape and bias parameters to the global best-fit values. The error bars thus reflect the observed scatter in the measured 11-band emissivities at a given redshift (Figure~\ref{fig:rest_spectrum}).
Our results indicate that, over cosmic time, $\Omega_{\rm dust}$ first increased by nearly an order of magnitude from $z=4$, reaching a broad peak at $z = 1$--1.5 to  $\sim10^{-5.5}$, before declining threefold to $\sim10^{-6}$ by the present. Note that since $\Omega_{\rm dust}$ is an accumulated quantity, the decline at late time indicates that the net change rate of dust mass turns negative, a trend we will discuss in more detail later.

\begin{table*}[t!]
\begin{center}
%\centering
\caption{\label{table:omega_dust_SFD_history} \normalsize Summary of Astrophysics Information---Cosmic Dust and Star Formation History}
\renewcommand{\arraystretch}{1.2}
\hspace{-1.6cm}
\begin{tabular}{rrrrrrrrr}
\hline 
$z$ & $\langle T \rangle$ & $T_{\rm peak}$ & log~$\Omega_{\rm{dust}}$ & log~$\Omega_{\rm{dust}}^{\rm{max}}$ & log~$\rho_{\rm IR}$  & log~$\dot{\rho}_{\star, \rm{IR}}$ & log~$\dot{\rho}_{\star, \rm{IR+UV}}$ & $f_{\rm obs}$\\ 
 & [K] & [K] & & & [erg~s$^{-1}$~Mpc$^{-3}$] & [M$_{\odot}$~yr$^{-1}~$Mpc$^{-3}$] & [M$_{\odot}$~yr$^{-1}~$Mpc$^{-3}$] & \\
 \hline 

0.10  &  $16.58^{+0.51}_{-0.57}$  &  $44.89^{+0.53}_{-0.61}$  &  $-5.86^{+0.07}_{-0.08}$  &  $-5.45$ &  $41.64^{+0.07}_{-0.08}$   &  $-1.92^{+0.07}_{-0.08}$  &  $-1.82^{+0.06}_{-0.07}$  &  $0.81^{+0.06}_{-0.06}$ \\
0.22  &  $16.31^{+0.36}_{-0.39}$  &  $44.42^{+0.36}_{-0.41}$  &  $-5.88^{+0.04}_{-0.05}$  &  $-5.45$ &  $41.68^{+0.04}_{-0.05}$   &  $-1.88^{+0.04}_{-0.05}$  &  $-1.77^{+0.04}_{-0.04}$  &  $0.78^{+0.04}_{-0.04}$ \\
0.34  &  $17.21^{+0.26}_{-0.28}$  &  $45.14^{+0.25}_{-0.31}$  &  $-5.75^{+0.03}_{-0.04}$  &  $-5.44$ &  $41.86^{+0.03}_{-0.04}$   &  $-1.69^{+0.03}_{-0.04}$  &  $-1.60^{+0.03}_{-0.03}$  &  $0.81^{+0.03}_{-0.03}$ \\
0.48  &  $17.86^{+0.21}_{-0.20}$  &  $45.59^{+0.20}_{-0.21}$  &  $-5.67^{+0.02}_{-0.03}$  &  $-5.43$ &  $41.99^{+0.02}_{-0.03}$   &  $-1.56^{+0.02}_{-0.03}$  &  $-1.48^{+0.02}_{-0.02}$  &  $0.83^{+0.03}_{-0.03}$ \\
0.64  &  $18.21^{+0.20}_{-0.22}$  &  $45.71^{+0.20}_{-0.23}$  &  $-5.63^{+0.02}_{-0.03}$  &  $-5.40$ &  $42.09^{+0.02}_{-0.03}$   &  $-1.46^{+0.02}_{-0.03}$  &  $-1.39^{+0.02}_{-0.02}$  &  $0.85^{+0.03}_{-0.02}$ \\
0.81  &  $19.32^{+0.29}_{-0.30}$  &  $46.57^{+0.26}_{-0.29}$  &  $-5.52^{+0.03}_{-0.03}$  &  $-5.38$ &  $42.26^{+0.03}_{-0.03}$   &  $-1.30^{+0.03}_{-0.03}$  &  $-1.24^{+0.03}_{-0.03}$  &  $0.87^{+0.02}_{-0.02}$ \\
1.00  &  $19.49^{+0.33}_{-0.33}$  &  $46.48^{+0.32}_{-0.32}$  &  $-5.53^{+0.04}_{-0.04}$  &  $-5.35$ &  $42.30^{+0.04}_{-0.04}$   &  $-1.25^{+0.04}_{-0.04}$  &  $-1.19^{+0.03}_{-0.04}$  &  $0.87^{+0.03}_{-0.03}$ \\
1.20  &  $19.42^{+0.35}_{-0.37}$  &  $46.20^{+0.35}_{-0.34}$  &  $-5.52^{+0.04}_{-0.04}$  &  $-5.34$ &  $42.36^{+0.04}_{-0.04}$   &  $-1.19^{+0.04}_{-0.04}$  &  $-1.13^{+0.04}_{-0.04}$  &  $0.86^{+0.03}_{-0.03}$ \\
1.43  &  $19.39^{+0.32}_{-0.32}$  &  $45.94^{+0.29}_{-0.29}$  &  $-5.57^{+0.03}_{-0.04}$  &  $-5.35$ &  $42.37^{+0.03}_{-0.04}$   &  $-1.19^{+0.03}_{-0.04}$  &  $-1.11^{+0.03}_{-0.03}$  &  $0.84^{+0.03}_{-0.03}$ \\
1.68  &  $20.69^{+0.29}_{-0.28}$  &  $46.86^{+0.26}_{-0.23}$  &  $-5.51^{+0.02}_{-0.02}$  &  $-5.39$ &  $42.48^{+0.02}_{-0.02}$   &  $-1.07^{+0.02}_{-0.02}$  &  $-1.00^{+0.03}_{-0.02}$  &  $0.84^{+0.03}_{-0.03}$ \\
1.96  &  $21.11^{+0.32}_{-0.35}$  &  $47.01^{+0.26}_{-0.32}$  &  $-5.58^{+0.02}_{-0.03}$  &  $-5.47$ &  $42.47^{+0.02}_{-0.03}$   &  $-1.09^{+0.02}_{-0.03}$  &  $-1.00^{+0.03}_{-0.03}$  &  $0.81^{+0.04}_{-0.03}$ \\
2.27  &  $22.50^{+0.33}_{-0.32}$  &  $47.93^{+0.30}_{-0.27}$  &  $-5.66^{+0.02}_{-0.02}$  &  $-5.60$ &  $42.44^{+0.02}_{-0.02}$   &  $-1.12^{+0.02}_{-0.02}$  &  $-1.01^{+0.03}_{-0.03}$  &  $0.78^{+0.05}_{-0.04}$ \\
2.61  &  $22.18^{+0.46}_{-0.53}$  &  $47.42^{+0.39}_{-0.45}$  &  $-5.88^{+0.03}_{-0.03}$  &  $-5.75$ &  $42.27^{+0.03}_{-0.03}$   &  $-1.28^{+0.03}_{-0.03}$  &  $-1.13^{+0.04}_{-0.04}$  &  $0.70^{+0.07}_{-0.06}$ \\
2.98  &  $21.89^{+0.66}_{-0.70}$  &  $46.92^{+0.56}_{-0.58}$  &  $-6.08^{+0.04}_{-0.05}$  &  $-5.92$ &  $42.13^{+0.04}_{-0.05}$   &  $-1.43^{+0.04}_{-0.05}$  &  $-1.23^{+0.06}_{-0.05}$  &  $0.64^{+0.08}_{-0.07}$ \\
3.40  &  $22.60^{+1.14}_{-1.25}$  &  $47.24^{+0.92}_{-1.02}$  &  $-6.28^{+0.06}_{-0.07}$  &  $-6.10$ &  $41.98^{+0.06}_{-0.07}$   &  $-1.58^{+0.06}_{-0.07}$  &  $-1.35^{+0.07}_{-0.07}$  &  $0.59^{+0.09}_{-0.08}$ \\
3.85  &  $24.07^{+1.84}_{-2.13}$  &  $48.15^{+1.38}_{-1.69}$  &  $-6.37^{+0.10}_{-0.13}$  &  $-6.29$ &  $41.95^{+0.10}_{-0.13}$   &  $-1.61^{+0.10}_{-0.13}$  &  $-1.39^{+0.08}_{-0.09}$  &  $0.61^{+0.09}_{-0.10}$ \\
\hline
\vspace{-0.3cm}\\
\multicolumn{9}{l}{ \parbox{0.96\textwidth}{
$z$---redshift; $\langle T \rangle$---mass-weighted mean dust temperature (Figure~\ref{fig:T_dust}); $T_{\rm peak}$---light-weighted dust temperature at peak emissivity $\epsilon_{\nu}$ (Figure~\ref{fig:T_dust}); $\Omega_{\rm{dust}}$---cosmic dust mass density parameter, fiducial result with warm-cold dust (Figure~\ref{fig:Omega_dust}); $\Omega_{\rm{dust}}^{\rm{max}}$---absolute upper limit (95\%) for the total warm, cold, very cold dust combined if allowing temperature bimodality for the coldest component (Figure~\ref{fig:Omega_dust}); $\rho_{\rm IR}$--- total infrared luminosity density integrating over 8--1000$~\mu$m (constrained by data down to 37.5~$\mu$m to exclude AGN contribution; Figure~\ref{fig:CSFR}); $\dot{\rho}_{\star, \rm{IR}}$---dust obscured cosmic star formation rate density in the CIB (Figure~\ref{fig:CSFR}); $\dot{\rho}_{\star, \rm{IR+UV}}$---total cosmic star formation rate density in the CIB plus UV (Figure~\ref{fig:CSFR}); $f_{\rm obs}$---dust obscured fraction of cosmic star formation (Figure~\ref{fig:CSFR}, upper panel).
}}\\
\end{tabular}
\end{center}
%\vspace{0.05cm}
\end{table*}

We compare our measurement to other emission-based $\Omega_{\rm dust}$ estimates in the literature, as shown in Figure~\ref{fig:Omega_dust}. These include \citet{2013ApJ...768...58T} in the yellow band, who used the projected power spectrum between the diffuse far-infrared and near-infrared backgrounds combined with a halo model for redshift evolution, and \citet{2011MNRAS.417.1510D}, \citet{2018MNRAS.475.2891D}, \citet{2020ApJ...892...66M}, \citet{2020MNRAS.491.5073P}, \citet{2021MNRAS.500..942D}, \citet{2024MNRAS.529.1130E}, \citet{2024A&A...690A..84T}, and \citet{2025A&A...696A.193B} in green symbols by integrating dust mass functions of objects from far-infrared galaxy surveys. In these studies, the absolute normalization of $\Omega_{\rm dust}$ scales inversely with the assumed opacity anchor $\kappa_0$. Our analysis, along with several others, adopts $\kappa_{850\,\mu{\rm m}} = 0.77~\rm cm^2\,g^{-1}$ from \citet{2000MNRAS.315..115D}, a commonly used value due to its implementation in one of the popular SED fitting codes \textsc{MAGPHYS} \citep{2008MNRAS.388.1595D}. However, this is by no means a consensus: values differing by a factor of two are often adopted for $\Omega_{\rm dust}$ estimates, and variations by an order of magnitude have been reported in broader contexts (see Appendix~C of \citealt{2025A&A...696A.193B} and Figure~1 of \citealt{2019MNRAS.489.5256C}). For consistency, we rescale all far-infrared based $\Omega_{\rm dust}$ to a common $\kappa_{850\,\mu{\rm m}} = 0.77~\rm cm^2\,g^{-1}$, except one.\footnote{\citet{2020MNRAS.491.5073P} adopt a $\kappa_0$ differing by a factor of two, but other assumptions compensate, yielding galaxy dust masses consistent with those fitted by \textsc{MAGPHYS} (same $\kappa_0$ as ours), so we do not rescale their result. This also implies a factor-of-two systematics even after homogenizing $\kappa_0$.}

There is substantial scatter among galaxy survey results beyond their statistical errors as shown in the plot, likely arising from cosmic variance and systematics in spectral sampling, SED fitting, and completeness corrections. The late-time decline in $\Omega_{\rm dust}$ is evident in \citet{2018MNRAS.475.2891D} and \citet{2011MNRAS.417.1510D}, in addition to ours. Overall, both \cite{2013ApJ...768...58T} and our CIB-based estimates trace the upper envelope of galaxy-based values. Since the redshift integral of our $\Omega_{\rm dust}$ is anchored to the projected FIRAS-\textit{Planck} CIB monopole at low frequencies \citep{2019ApJ...877...40O}, if correct, galaxy surveys reporting consistently lower $\Omega_{\rm dust}$ would underpredict the CIB. Although these object-based estimates require faint-end extrapolation, they might not necessarily miss CGM dust around detected galaxies given that the main far-infrared data from {\it Herschel} lack the resolution to isolate galactic disks soon beyond $z=0$. This complexity, along with related concerns, is naturally circumvented in our intensity mapping approach, which offers several advantages: it captures all emission associated with the LSS without requiring extrapolation to faint galaxies or diffuse components, features spectral coverage that spans the Rayleigh-Jeans tail at all redshifts for temperature-resilient dust mass estimates, and benefits from a large sky coverage of $\sim 1\pi$~steradian rendering the cosmic variance negligible.

Figure~\ref{fig:Omega_dust} also presents different types of cosmic dust density estimates: a more theoretical one by \cite{2004ApJ...616..643F} using the depletion of Silicon and Carbon together with the metal yield given by stellar mass density at $z=0$, an estimate of dust in the extended halos of galaxies from the galaxy-reddening correlation function by \cite{2010MNRAS.405.1025M}, another estimate of CGM dust from reddening associated with Mg II absorbers by \cite{2012ApJ...754..116M}, and finally through metal depletion in HI damped Ly$\alpha$ absorbers by \cite{2020ARA&A..58..363P}. These absorption-based results estimate only partial contributions to $\Omega_{\rm dust}$ excluding at least dense parts of galactic disks. This is due to both the nature of various absorber tracers and the fact that optically selected QSOs are biased against lines of sight intercepting dusty disks. The corresponding estimates are thus plotted as lower limits. Calibrating extinction-based dust mass requires an extinction-to-mass conversion, typically derived from dust grain size distribution models. As with emission, such coefficients are uncertain and can vary substantially depending on the dust type. For example, Milky Way type dust is about twice as massive as Small Magellanic Cloud type at a given visible extinction \citep{2001ApJ...548..296W}. Comparing extinction- and infrared-emission-based results thus relies on two poorly constrained conversion factors. Nonetheless, to facilitate discussion, in Figure~\ref{fig:Omega_dust}, we follow \cite{2025A&A...696A.193B} and rescale these optical reddening estimates to be compatible with far-infrared ones anchored at $\kappa_{850\,\mu{\rm m}} = 0.77~\rm cm^2\,g^{-1}$.\footnote{The rescaling factors are 0.83 for \cite{2010MNRAS.405.1025M} and \cite{2012ApJ...754..116M} via \cite{2003ARA&A..41..241D}, and 0.61 for \cite{2020ARA&A..58..363P}.} While these reddening-based $\Omega_{\rm dust}$ are lower limits, it is reassuring that they are of the same order as emission-based estimates. More precise comparisons are likely to inform the extinction-to-mass and/or emission-to-mass ratios, as well as incompleteness, rather than the absolute dust mass.

%\paragraph{Late time decline}
We now examine the robustness of the late-time $\Omega_{\rm dust}$ decline we find. First, the same three-fold drop between $z=1$ and the present could be found if we simply interpolate the measured CIB emissivity in the Rayleigh-Jeans side at, e.g., $\sim 400$~GHz, in Figure~\ref{fig:rest_spectrum} and correct for the best-fit bias evolution $b(z)$ from Figure~\ref{fig:bias}. This reassures that the information on dust mass is nearly model-independent in terms of choices in SED parameterization, as expected. Next, we examine the impact of the bias evolution $b(z)$, our main systematic uncertainty. Empirically, the raw clustering redshift amplitudes (Figure~\ref{fig:dIdzb}) at the Rayleigh-Jeans tail constrain the product $\Omega_{\rm dust}\,b$, which we find to increase by a factor of 6 from $z=0$ to $z=1$. If $\Omega_{\rm dust}$ were to remain constant over this redshift range, the CIB bias $b$ would need to increase by the same factor, implying $b(z=1) \sim 6$, which is firmly ruled out when incorporating the FIRAS-{\it Planck} CIB monopole into our data vector. Conversely, if $b$ evolves more slowly than our fiducial posterior in Figure~\ref{fig:bias}, then $\Omega_{\rm dust}$ must decline even more steeply at $z<1$. We therefore conclude that the late-time drop in $\Omega_{\rm dust}$ is a robust feature inferred from the CIB data, i.e., not driven by the functional form in Equation~\ref{eq:z_evo_norm}.

The $\Omega_{\rm dust}$ decline at $z<1$ suggests that the total dust destruction rate exceeds the production rate at late times. Destruction mechanisms include astration, supernova shocks, shattering, and sputtering, while dust production is primarily driven by grain growth in the ISM following initial yields from supernovae and asymptotic giant branch stars \citep[e.g.,][]{1998ApJ...501..643D,2009MNRAS.394.1061H,2018MNRAS.478.4905A}.  

Theoretical studies do not always reproduce the late-time decline of $\Omega_{\rm dust}$, as many of the aforementioned sink and source terms remain uncertain and could individually be as large as the total net sum \citep[see Figure~9 in][]{2020MNRAS.493.2490T}. Even when the decline is reproduced, the underlying driving mechanisms may differ \citep[e.g.,][]{2021MNRAS.503.4537F,2023MNRAS.521.6105P}.  

Compared to gas, \cite{2020ApJ...902..111W} found that the molecular hydrogen $\rm H_2$ density in the universe declines by a factor of $\sim6$ from $z=1$--1.5 to $z=0$, while the atomic HI density in galaxies shows little to no evolution. Explaining this balance requires a highly dynamic picture involving cosmological inflow and gas consumption through star formation. Combining their total $\rm HI + H_2$ gas density in galaxies (excluding ionized gas, which is predominantly in the CGM and IGM) with our dust density estimate, we find that the cosmic dust-to-gas ratio remains nearly constant, with a possible mild decline from $0.3\%$ at $z = 1$--1.5 to $0.2\%$ at $z = 0$.

In discussing the global dust budget, one highly relevant process is the expulsion of dust from galaxies into the CGM via stellar and/or AGN-driven winds, followed by destruction through thermal sputtering. Sputtering can be highly efficient when dust is mixed with the million-Kelvin hot halo gas \citep{1979ApJ...231..438D}. Nonetheless, a fraction of the dust survives, as suggested by the halo-scale galaxy-reddening correlation detected in \cite{2010MNRAS.405.1025M}, \cite{2015ApJ...813....7P}, and \cite{2025arXiv250304098M}. Far-infrared dust emission is detected in the CGM, so far, in only six local dwarf galaxies in \cite{2018MNRAS.477..699M}, with a $\sim 20$\% dust mass fraction outside the stellar disks. More recently, JWST detected PAH plumes ejected into the inner CGM of NGC 891 \citep{2024A&A...690A.348C}, demonstrating the survival of small grains in hot halos, while it is unclear if they can be long-lived. Simulations have begun to reveal mechanisms that allow dust to persist in the multiphase CGM, particularly within cold clouds \citep{2021MNRAS.503..336K,2024ApJ...974...81R}. Dust mass might even grow in situ in these clouds by accreting gas-phase metals \citep{2024MNRAS.528.5008O}. On further larger scales, metals are known to exist in the diffuse IGM \citep{1996AJ....112..335S,2006MNRAS.373.1265O}, though whether dusty outflows can reach the IGM remains uncertain.  
Could outflows be responsible for the $\Omega_{\rm dust}$ decline observed at $z<1$? Both the one-zone model of \cite{2024MNRAS.528.5008O} and the cosmological hydrodynamic simulation of \cite{2019MNRAS.490.1425L} suggest that the initial dust mass budget entrained in outflows can be substantial, potentially exceeding 50\% of the total dust ever produced in the ISM \citep{2011arXiv1103.4191F}. However, the net efficiency of CGM dust sputtering remains highly uncertain. To gain deeper physical insight, we can further explore constraints from the dust temperature information.

\begin{figure*}[ht]
\centering
\includegraphics[width=0.765\textwidth]{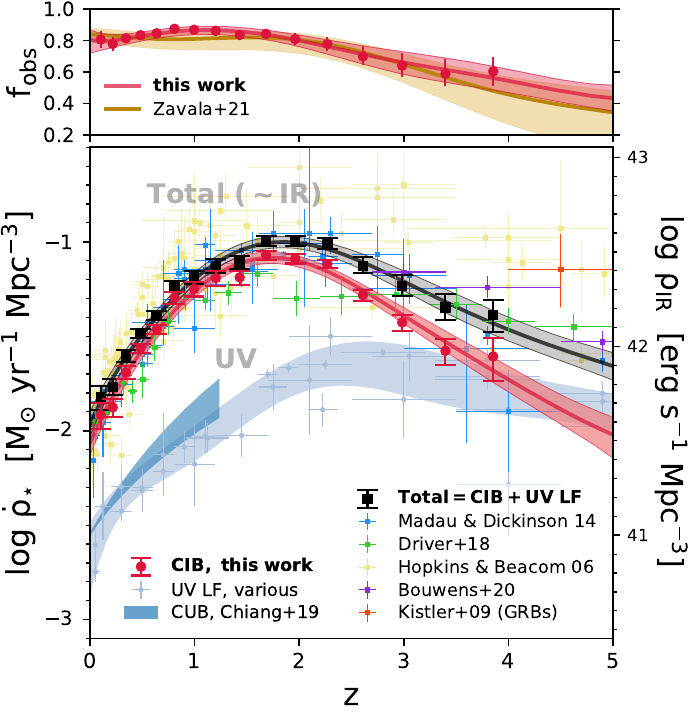}
\caption{Cosmic star formation history. Red band and data points show our posterior dust-obscured star formation rate density from CIB tomography, with the raw total infrared luminosity density labeled on the right. For the total star formation (black band and data points), we add the CIB to a compilation of unobscured UV contributions from literature luminosity function integrals without dust correction (light steel blue; see text for references). The upper panel quantifies the obscured fraction $f_{\rm obs}$, given by the ratio of IR to total star formation, indicating a heavily dust-obscured star formation history. Literature measurements are overlaid as labeled, with all values converted to assume a \cite{2003PASP..115..763C} IMF. Our CIB tomography is robust against systematics common in galaxy surveys, such as faint-end incompleteness, photometric redshift errors, and large SED extrapolations, and has negligible cosmic variance, utilizing $1\pi$ sky coverage instead of pencil-beam surveys. As with $\Omega_{\rm dust}$, our star formation normalization is empirically set by the CIB monopole and its associated uncertainties. The resulting error at a given redshift is dominated by the detailed bias evolution (Figure~\ref{fig:bias}), with photon noise already subdominant.}
\label{fig:CSFR}
\end{figure*}

If CGM dust survives the hot halo environment within cold clouds and shares similar temperatures with ISM dust, our CIB analyses cannot distinguish between them. On the other hand, a large amount of hot CGM dust is already ruled out, as its far-infrared emission, though extended, would be bright and difficult to hide in observations of nearby galaxies. What we can test, instead, is a third scenario: an additional contribution of very cold dust to the overall dust mass budget. As already mentioned, using the extended model presented in section~\ref{sec:cold_CGM} and MCMC run in Section~\ref{sec:MCMC}, very cold dust is not significantly detected and only leads to upper limits. This limit is less stringent at lower redshifts, primarily due to the adopted temperature evolution (Equation~\ref{eq:T_cold}) coupled with the inherently low emissivity of colder dust. By adding the 95\% upper limit of the very cold component to the $2\sigma$ upper bound of the detected warm-cold component, we derive an upper limit for the maximum $\Omega_{\rm{dust}}$ allowed by the CIB, shown as the black dotted line in Figure~\ref{fig:Omega_dust}. Indeed, none of the galaxy survey results (green symbols) exceeds this limit significantly, supporting consistency with our CIB constraint.

\subsection{Cosmic Star Formation}\label{sec:CSFR} 

The total energy in the CIB serves as a powerful probe of cosmic star formation, as it originates from UV and visible light emitted by stars, absorbed by dust, and re-emitted in the infrared. We quantify cosmic star formation using the comoving star formation rate density, $\dot{\rho}_{\star}$, in $\rm M_{\odot}\ yr^{-1}\ Mpc^{-3}$ \citep{1996ApJ...460L...1L,1996MNRAS.283.1388M}. A corresponding density parameter can be defined as $\dot{\Omega}_{\star} = \dot{\rho}_{\star}/\rho_{\rm crit}$, though we primarily report $\dot{\rho}_{\star}$ for more direct comparisons with literature.  

Figure~\ref{fig:CSFR} presents our posterior cosmic star formation history derived from CIB tomography in the red line, with the 68\% range indicated by the shaded region. This is obtained by computing $\rho_{\rm IR}$, the total comoving infrared luminosity density in the CIB, shown on the right axis, by integrating each posterior SED from the MCMC samples between 8 and 1000~$\rm \mu$m. The medians of these SEDs, weighted by the bias posterior, correspond to the lines in Figure~\ref{fig:rest_spectrum}, and the resulting $\rho_{\rm IR}$ integrals are agnostic to the model’s physical interpretation, as long as it provides a good fit to the data. After calculating the $\rho_{\rm IR}$ posteriors redshift by redshift, we find that the evolution can be well fitted using the functional form in Equation~\ref{eq:z_evo_norm}, yielding:
\begin{eqnarray}
\rho_{\rm IR}(z)\, =\, 3.3\times10^{41}\,\frac{(1+z)^{2.68}}{1+[(1+z)/2.93]^{6.93}}\,
\label{eq:rho_IR_fit}
\end{eqnarray}
in $\rm erg\ s^{-1}\ Mpc^{-3}$. The statistical uncertainty in $\rho_{\rm IR}(z)$ is about 0.04 dex (10\%), with dominant systematics likely from the redshift evolution of the CIB bias $b$, followed by external monopole constraints and photometric calibrations in {\it Planck} and {\it IRAS} at high frequencies. 

To calculate the star formation rate density, a conversion is then applied:
\begin{eqnarray}
\dot{\rho}_{\star} = {\rm K}_{\rm IR} \times \rho_{\rm IR}\ .
\label{eq:TIR_to_SFR}
\end{eqnarray}
We assume ${\rm K}_{\rm IR} = 2.795 \times 10^{-44}$~$\rm M_{\odot}\ yr^{-1}\ erg^{-1}\ s$, converted from the value originally given in \cite{1998ARA&A..36..189K} for a \cite{1955ApJ...121..161S} stellar initial mass function (IMF) to our preferred \cite{2003PASP..115..763C} IMF. For a non-parametric redshift evolution check, Figure~\ref{fig:CSFR} also presents per-redshift $\dot{\rho}_{\star}$ values as red data points, obtained from the MCMC run with one free amplitude ($\rho_{\rm d}$ in our model) per redshift bin, same run as that used for the per-redshift $\Omega_{\rm dust}$ in Figure~\ref{fig:Omega_dust}. Overall, we observe a pronounced peak at $z\sim2$, when the universe was most efficient at forming stars. 

Although we follow the convention of integrating over 8--1000~$\rm \mu$m for $\rho_{\rm IR}$, our CIB spectral fitting includes only data points down to 37.5~$\mu$m (8~THz). Below this wavelength, our posterior spectrum follows a featureless power-law. This spectral cut ensures high-fidelity exclusion of AGN contributions in the mid-infrared: \cite{2018MNRAS.478.4238D} estimated that even for sources selected as AGNs, their luminosity over 30--1000~$\rm \mu$m is consistent with being fully dominated by star formation. We thus expect negligible AGN contamination in the cosmic star formation rate densities reported here.  

Strictly speaking, our CIB-based $\dot{\rho}_{\star}$ represents only the dust-obscured contribution. In Figure~\ref{fig:CSFR}, we additionally show the unobscured $\dot{\rho}_{\star}$ derived from integrating literature UV luminosity functions of galaxies without correcting for dust extinction. These are displayed as light steel blue data points from the \cite{2014ARA&A..52..415M} compilation and more recent measurements from \cite{2015ApJ...803...34B}, \cite{2015ApJ...810...71F}, and \cite{2016MNRAS.456.3194P}, and we perform a spline fit shown in the smooth band. For these UV contributions to $\dot{\rho}_{\star}$, we assume a conversion factor following \cite{2020ApJ...902..112B}:
\begin{eqnarray}
\dot{\rho}_{\star} = {\rm K}_{\rm UV} \times \rho_{\rm UV},
\label{eq:UV_to_SFR}
\end{eqnarray}
with ${\rm K}_{\rm UV} = 7.143 \times 10^{-29}$~$\rm M_{\odot}\ yr^{-1}\ erg^{-1}\ s^{-1}\ Hz^{-1}$ for the \cite{2003PASP..115..763C} IMF, where $\rho_{\rm UV}$ represents the integrated specific UV comoving luminosity density at rest-frame $1500~\rm \AA$.  

For comparison, we also show, in the darker blue band, the intensity mapping tomography result for the UV background from \cite{2019ApJ...877..150C}. This measurement is based on cross-correlating diffuse UV photons from GALEX with reference redshifts in SDSS using a methodology similar to this work. The resulting UV background $\dot{\rho}_{\star}$ is slightly higher but remains consistent with integrated galaxy light.

To derive the total star formation history, we sum the contributions from the CIB and UV and present the result as the black band and data points in Figure~\ref{fig:CSFR}. For the UV, the compilation of galaxy luminosity integral is used instead of the diffuse UV background tomography from \cite{2019ApJ...877..150C} as the latter does not cover high redshifts where rest-frame UV has shifted out of the GALEX bands. The best-fitting function for our total CIB plus UV cosmic star formation history, using the form in Equation~\ref{eq:z_evo_norm}, is:
\begin{eqnarray}
\dot{\rho}_{\star}(z)\, =\, 0.011\,\frac{(1+z)^{2.64}}{1+[(1+z)/2.95]^{5.94}}\,
\label{eq:csfd_total_fit}
\end{eqnarray}
in $\rm{M_{\odot}\ yr^{-1}\ Mpc^{-3}}$. 

We find that the CIB dominates over the UV over an extended period. In the upper panel of Figure~\ref{fig:CSFR}, we show the obscured fraction $\rm f_{obs}$ by dividing the CIB $\dot{\rho}_{\star}$ by the total from CIB plus UV. Our results indicate that cosmic star formation occurs in a remarkably dust-obscured mode, with $\rm f_{obs}$ exceeding 80\% for more than 10 Gyr between $z=2$ and the present. We find that $\rm f_{obs}$ declines gradually toward higher redshifts but remains at approximately 60\% at $z=4$. This suggests that dust enrichment and obscuration in the universe are highly efficient and spatially co-located with young stars on galactic scales. Our $\rm f_{obs}$ estimates are generally consistent with those from \cite{2021ApJ...909..165Z} for resolved galaxies (light brown band in Figure~\ref{fig:CSFR}), though their uncertainties are substantial, especially at high redshifts, due to the small number of galaxies available for the estimation and the uncertainty in completeness correction. 

Between $z = 1.5$ and the present, star formation declined by an order of magnitude---a factor larger than the threefold decline in accumulated dust mass in Figure~\ref{fig:Omega_dust}. As one is a rate and the other an integral, the comparison is not to be taken too far, but notably, the dust obscuration fraction, $\rm f_{obs}$, remains roughly constant over this period of active evolution in both star formation and dust mass.

There is a rich body of literature estimating the total cosmic star formation history using various approaches. In Figure~\ref{fig:CSFR}, we compare our results with the following compilations: \cite{2014ARA&A..52..415M} and \cite{2018MNRAS.475.2891D}, which are based on UV and far-infrared emission from dropout galaxies; \cite{2020ApJ...902..112B}, which corrects for the contribution of submillimeter galaxies missed in dropout-based measurements; the earlier \cite{2006ApJ...651..142H} collection, which combines dropout and H$\alpha$ measurements; and \cite{2009ApJ...705L.104K}, which estimates star formation history based on gamma-ray bursts at high redshifts. All results are converted to the \cite{2003PASP..115..763C} IMF, either by adopting the same ${\rm K}_{\rm IR}$ and ${\rm K}_{\rm UV}$ (Equations~\ref{eq:TIR_to_SFR} and \ref{eq:UV_to_SFR}) used in this work or by dividing the star formation rate densities originally derived under the \cite{1955ApJ...121..161S} IMF by a factor of 1.61. Our results are consistent with \cite{2014ARA&A..52..415M}, \cite{2020ApJ...902..112B}, and \cite{2018MNRAS.475.2891D} but are lower than those from \cite{2006ApJ...651..142H} and \cite{2009ApJ...705L.104K}. Not shown in the plot are other CIB  estimates from, e.g., \cite{2018A&A...614A..39M} and \cite{2024JCAP...05..058Y}, which are broadly consistent with ours.

Similar to $\Omega_{\rm dust}$, our CIB tomography-based cosmic star formation history has several advantages over previous measurements in the literature:  
(1) negligible cosmic variance given the use of $\sim1\pi$ steradian of sky, compared to sub-degree fields in some deep galaxy surveys,  
(2) selection-function-free and fully complete to the faintest galaxy populations,  
(3) precise redshift information obtained via cross-correlations with spectroscopic references,  
(4) superior frequency sampling of the far-infrared SED, enabling an empirical 8--1000~$\rm \mu$m integral rather than relying on a template extrapolation from a few data points (sometimes even just single-band) in some galaxy-based measurements.  

For these reasons, it is compelling to consider the tomographic CIB-based measurement as a new standard reference for the cosmic star formation history up to $z\sim 4$. In addition to summarizing a key component of the cosmic inventory, our results can serve as a benchmark for testing and validating cosmological simulations of galaxy formation \citep{2020NatRP...2...42V}. We summarize our CIB-based constraints on the total infrared luminosity density, cosmic star formation rate density, obscured fraction, and other dust-related quantities in Table~\ref{table:omega_dust_SFD_history}, and provide the full dataset as an electronic table at the URL in footnote~\ref{fn:datarelease}.

\section{Discussion} \label{sec:discussion}

\subsection{Data Fusion in Astronomy}

This work exemplifies large-scale data fusion in modern astronomy. Its integrative framework enables robust measurements of faint and diffuse structures in the background light, and supports accurate cosmological inferences by accounting for a comprehensive set of modeling uncertainties under maximally empirical constraints. Our analysis brings together multiple sources of data  and techniques along three main axes:

\begin{itemize}[leftmargin=*]
\item Tomography: We combine intensity mapping, which captures the aggregate emission of all sources but lacks redshift information, with galaxy surveys, which offer precise redshifts for discrete objects but are limited by flux thresholds. This fusion allows us to break the line-of-sight projection and tomographically reconstruct the evolution of the CIB. The synergy offers both completeness (capturing faint galaxies and diffuse emission) and redshift resolution, yielding robust, foreground-resistant measurements.

\item Spectral coverage for the dust SED: We utilize far-infrared intensity maps across 11 bands from \textit{Planck}, \textit{Herschel}, and \textit{IRAS}, spanning a factor of 50 in observed frequency (250 in rest-frame). This broad spectral range enables empirical dust mass and star-formation constraints, overcoming limitations in previous studies that relied on extrapolations from sparsely sampled or template dust SEDs.

\item Computation: Our analysis involves computing nearly one million spatial two-point correlation measurements (11 bands $\times$ 160 redshift bins $\times$ 500 bootstrap samples). The raw data itself is 
substantial---over 100 million sky pixels from the unmasked regions across 11 intensity maps, combined with 3 million spectroscopic redshifts from SDSS. 

\end{itemize}

Our data-intensive measurements are applied uniformly across 12.3~Gyr of cosmic time, avoiding redshift-stitching artifacts and enabling a seamless view of cosmic star formation and dust evolution. This methodology not only overcomes the limitations of individual datasets but also enhances signal detection, demonstrating how cross-survey synthesis can effectively extract cosmological and astrophysical information. Such an approach is poised to play an increasingly central role in future astrophysical analyses.

\begin{figure*}
\center
\includegraphics[width=1\textwidth]{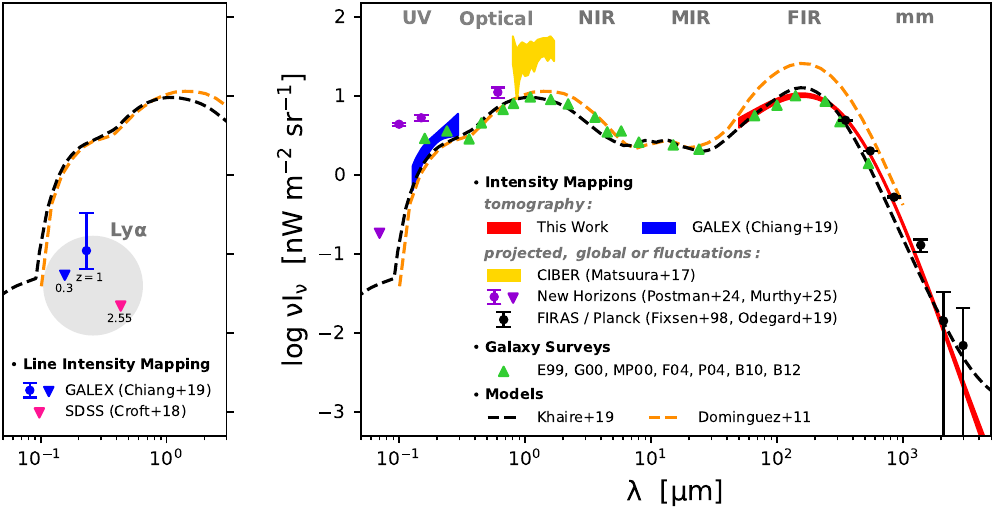}
\caption{Summary of EBL monopole intensity $\nu I_{\nu}$ from UV to millimeter wave. Intensity mapping results are organized into two categories. The first is cross-correlation-based tomography, the method most reliable against foregrounds and capable of capturing, if present, physically diffuse light in the LSS. This includes the thin red band for our CIB result (68\% range, same as Figure~\ref{fig:monopole}) and the blue band for the UV background tomography in \cite{2019ApJ...877..150C}. \cite{2019ApJ...877..150C} also obtained a detection (limit) of cosmic $\rm Ly\alpha$ at $z=1$ ($0.3$), which we show on the left panel in blue symbol, together with an upper limit of $\rm Ly\alpha$ at $z=2.55$ from \cite{2018MNRAS.481.1320C} in pink. The second category is intensity mapping using global intensity or projected auto-fluctuations with known foreground subtracted. This includes {\it CIBER} \citep{2017ApJ...839....7M}, {\it New Horizons} \citep{2024ApJ...972...95P,2025AJ....169..103M}, {\it COBE} FIRAS \citep{1998ApJ...508..123F}, and {\it Planck} \citep{2019ApJ...877...40O}. A selected compilation of integrated galaxy light measurements \citep{1999A&A...351L..37E,2000ApJ...542L..79G,2000MNRAS.312L...9M,2004ApJS..154...39F,2004ApJS..154...70P,2010A&A...518L..30B,2012A&A...542A..58B} is marked in green triangles as lower limits following \cite{2016RSOS....350555C} based on an earlier figure from \cite{2006A&A...451..417D}. We also show two synthesized models combining galaxy abundance, redshift distributions, and SEDs from \cite{2019MNRAS.484.4174K} and \cite{2011MNRAS.410.2556D} as thick dashed lines.}
\label{fig:EBL_monopole}
\end{figure*}

\subsection{New Compilation of EBL Monopole}\label{sec:EBL_monopole} 

The integrated sky intensity for the EBL across all wavelengths, or the full spectrum of the universe, is a fundamental summary of cosmological structure formation coupled with galaxy formation. Here, we discuss our results together with measurements in the literature. The multiband redshift tomography in this work gives a foreground-free measurement of the CIB monopole from 100~GHz to 5~THz (Figure~\ref{fig:monopole}). Since constraints from the FIRAS-informed {\it Planck} monopole up to 545~GHz is used, our result effectively extrapolates it by one order of magnitude into shorter wavelength where the Milky Way foreground is much stronger, which precludes a robust estimate before.

Figure~\ref{fig:EBL_monopole} shows a new compilation of EBL monopole intensity $\nu I_{\nu}$ in $\rm nW\,m^{-2}\,sr^{-1}$ from the UV to the millimeter. We organize observational results into two complementary categories. First, counting individually identified objects in galaxy surveys provides robust lower bounds on the EBL monopole. Following the review in \cite{2016RSOS....350555C}, we plot measurements from \cite{1999A&A...351L..37E}, \cite{2000ApJ...542L..79G}, \cite{2000MNRAS.312L...9M}, \cite{2004ApJS..154...39F}, \cite{2004ApJS..154...70P}, \cite{2010A&A...518L..30B}, and \cite{2012A&A...542A..58B} in green triangles. A more recent compilation is given in \cite{2018ApSpe..72..663H}, while the results have basically converged over the wavelengths considered, as the integrated galaxy light measurements do not require redshift estimates. These galaxy surveys are free of diffuse foreground and can be cleaned for point source foreground (i.e., stars) effectively at high latitudes. However, contributions from faint galaxies, diffuse intra-halo light (including halo stars, CGM gas, and dust), and potential IGM emission could be missing in some wavelengths. For comparison, we also show two models from \cite{2011MNRAS.410.2556D} and \cite{2019MNRAS.484.4174K}. In the wavelength range considered here above ionizing UV, they are constructed mostly via synthesizing galaxy number count, redshift distributions, and SEDs. These models are thus, by construction, quite consistent with integrated galaxy light measurements in green symbols. 

The second set of monopole measurements aims for a more complete EBL census using the emerging technique of intensity mapping, which we organize into two sub-categories in Figure~\ref{fig:EBL_monopole}. The first is projected global intensity or fluctuations through auto-correlations from FIRAS \citep{1998ApJ...508..123F}, FIRAS plus {\it Planck} \citep{2019ApJ...877...40O}, {\it CIBER} (\citealt{2017ApJ...839....7M}, see also \citealt{2014Sci...346..732Z}), and the recent  {\it New\ Horizons} results \citep{2024ApJ...972...95P,2025AJ....169..103M}. This method is more direct, while due to the projected nature, the measured raw amplitudes also respond to foregrounds directly. Substantial efforts have been invested in understanding and removing foregrounds of both instrumental and astrophysical origins. Some of these results are significantly higher than those from galaxy surveys, especially in the UV and near-infrared, hinting at potential missing light in the EBL not counted in galaxy surveys. However, it is still possible that there might be unidentified foregrounds yet to be removed. The second sub-category of intensity mapping is cross-correlation-based redshift tomography. This is, so far, applied in the far-infrared using 11 {\it Planck}, {\it Herschel}, and {\it IRAS} bands in this work (red thin band) and in the UV from \cite{2019ApJ...877..150C} cross-correlating diffuse light in GALEX (blue) with SDSS redshifts. This tomographic approach, as demonstrated in this work, is insensitive to foregrounds as the reference galaxies used are already extragalactic and do not correlate with emission within the Milky Way or the solar system. Meanwhile, intensity tomography is not limited by surface-brightness thresholds in the LSS. Encouragingly, they give quite consistent results compared to integrated galaxy light, lowering the allowed budget for potential diffuse CGM and IGM emission. Unfortunately, this method currently does not allow us to isolate the EBL from the Epoch of Reionization (EoR) in the early universe due to the lack of a large galaxy reference sample. To probe the EoR, future studies could consider cross-correlating two line intensity tracers. 

Interestingly, cross-correlation-based intensity tomography can already extract line information using even broadband data, as a line would enter and leave a given band at different redshifts. In the left panel, we show constraints for Ly$\alpha$ emission in blue symbols for the marginal detection at $z=1$ and limit at $z=0.3$ from \cite{2019ApJ...877..150C} using diffuse light in GALEX, and the pink limit at effectively $z=2.55$ from \cite{2018MNRAS.481.1320C} using SDSS fiber spectra cross Ly$\alpha$ forest as reference matter tracer. These Ly$
\alpha$ constraints are not very stringent yet only because the data used are from surveys not designed for intensity mapping. Some attempts for intensity-galaxy cross-correlations have also been made to probe lines in the far-infrared to millimeter \citep[e.g.,][]{2024arXiv240607861R}. With more dedicated experiments or more suitable datasets, future constraints could be promising. These include, e.g., HETDEX for Ly$\alpha$ \citep{2023ApJ...958....4L}, SPHEREx for H$\alpha$, H$\beta$, [OII], and [OIII] \citep{2022ApJ...925..136C}, and about a dozen submillimeter experiments for CO and [CII].\footnote{\url{https://lambda.gsfc.nasa.gov/product/expt/lim_experiments.html}} Lastly, we remark that in the coming era of line-intensity mapping experiments, in addition to strong lines, the same intensity data cubes should also be used to push our understanding of the EBL continuum, and the exact tomographic cross-correlations method in this work could be used.

\section{Summary}\label{sec:summary}

The CIB encapsulates the total thermal dust emission in the universe, tracing cosmic star formation driven by gravitational structure growth. For precision cosmology and astrophysics, however, we need to directly measure the evolving CIB spectrum beyond simplistic modeling assumptions or SED templates often used in the literature. The key challenge lies in the lack of redshift information as well as the superposition of the strong foreground of Galactic dust.

We probe the CIB spectrum as a function of redshift over $z=0$--4 by measuring an extensive set of tomographic CIB-galaxy cross-correlations in the two-halo regime. We use 11 far-infrared intensity maps from {\it Planck}, {\it Herschel}, and {\it  IRAS} over 100--5000~GHz, with effective foreground mitigation using CSFD, the CIB-free Milky Way dust template from \cite{2023ApJ...958..118C}. We cross-correlate diffuse far-infrared photons in each band with the positions of 3 million spectroscopic reference galaxies and QSOs from SDSS, BOSS, and eBOSS in tomographic redshift bins. Correcting these correlation amplitudes with reference galaxy bias and matter clustering, we probe the bias-weighted redshift distribution $b(dI/dz)$ for the CIB in each band up to $z=4$ (Figure~\ref{fig:dIdzb}). 

The 11-band CIB redshift distributions collectively trace a single evolving spectrum, quantified as the bias-weighted mean CIB emissivity $\epsilon_{\nu} b$ (Figure~\ref{fig:rest_spectrum}). This tomographic spectrum is sampled at 176 frequencies in the rest frame (11 bands $\times$ 16 redshift bins or $1+z$ shifts), making it more finely sampled than any individual dusty galaxy SED reported in the literature.

We break the bias-emissivity degeneracy by adding external CIB monopole constraints from  FIRAS and {\it Planck}. This allows us to determine the normalization of the tomographic spectrum empirically. Doing so also reveals that two out of the three FIRAS calibrations from \cite{1998ApJ...508..123F} have high-frequency foreground under-subtracted, and only that cleaned by HI data is compatible with our result (Figure~\ref{fig:monopole}). 

To interpret our tomographic CIB spectrum, we introduce an ``ensemble cosmic dust'' model, which generalizes the commonly used single-temperature graybody spectrum to one with a flexible distribution of temperatures. By fitting it to our evolving CIB spectrum, we obtain a smooth interpolation of the already well-sampled data and extract astrophysical information. We find a broad dust temperature distribution, reflecting the diverse galaxy populations and local environments hosting cosmic dust. The mass-weighted mean temperature, significantly lower than light-weighted ones, increases from $16$~K at $z = 0$ to $25$~K at $z=4$, indicating more intense interstellar heating at higher redshifts. Our CIB-based dust temperature evolves more gradually than that reported for bright dusty galaxies, indicating possible selection effects for these highly biased objects compared to the cosmic mean.

Using low-frequency CIB amplitudes, we constrain $\Omega_{\rm dust}$, the cosmic dust mass density parameter (Figure~\ref{fig:Omega_dust}). Over 12 Gyr of cosmic history, $\Omega_{\rm dust}$ first increases tenfold from $z=4$ to reach a broad peak at $z\sim 1$--$2$. Below $z=1$, $\Omega_{\rm dust}$ declines by a factor of 3, indicating more rapid dust destruction than production at late times. Our mass estimate is complete for all infrared photons tracing the LSS. In contrast to galaxy surveys, it does not depend on the detection of sources above a given surface brightness threshold.

We integrate our tomographic CIB spectrum to derive the total infrared luminosity density up to $z=4$, achieving constraints better than 0.04 dex across redshifts. This precision is enabled by the full sampling of the far-infrared SED and the large $\sim$1$\pi$ steradian sky coverage of our data, suppressing cosmic variance to a negligible level. Assuming a conversion factor, we trace the cosmic star formation history over 90\% of cosmic time (Figure~\ref{fig:CSFR}). Our intensity-mapping-based star formation history is broadly consistent with that from galaxy surveys \citep[e.g.,][]{2014ARA&A..52..415M}, while offering higher precision, requiring no completeness correction, and being free from photometric redshift uncertainties. Comparing our CIB result with UV luminosity function studies, we find that cosmic star formation is 80--90\% dominated by the dust-obscured mode at $z<2$, and the obscured fraction remains substantial at 60\% at $z=4$.

This study provides a precision measurement of the evolving mean CIB spectrum over $0 < z < 4$, enabling a comprehensive census of cosmic dust and star formation and revealing a heavily dust-obscured history of galaxy formation. By combining our new CIB constraints with multi-wavelength background light measurements from the literature, we compile updated panchromatic EBL monopole amplitudes (Figure~\ref{fig:EBL_monopole}). Methodologically, this work demonstrates the power of large-scale data fusion, highlighting the growing importance of cross-survey synergy in astronomy. Our CIB results will support a range of CMB experiments, enable the use of the CIB as a tracer of LSS for precision cosmology, and inform theoretical modeling of dust in and out of galaxies. To facilitate future studies, we release our main products—the tomographic CIB spectrum and redshift distributions—to the community (URL in footnote~\ref{fn:datarelease}).

\begin{acknowledgments}
Y.C. acknowledges the support of the National Science and Technology Council of Taiwan through grant NSTC 111-2112-M-001-090-MY3 and Academia Sinica through the Career Development Award AS-CDA-113-M01. We thank Eiichiro Komatsu for early discussions that helped motivate this work on CIB tomography; Andrew Hopkins and John Beacom for insights into systematic treatments of cosmic star formation estimates; Heidi Wu for discussions on CIB bias; Marco Viero, Michael Zemcov, and D\'eborah Paradis for suggestions on handling {\it Herschel} data; Bovornpratch Vijarnwannaluk for discussions on excluding AGN contributions; and the anonymous referee for helpful comments. We acknowledge the use of data provided by the CADE, a service of IRAP-UPS/CNRS. 

\end{acknowledgments}

\bibliography{references}{}
\bibliographystyle{aasjournal}

\end{document}